\documentclass[aps,prd,superscriptaddress,showpacs,nofootinbib,floatfix,twocolumn]{revtex4}

\usepackage{graphicx}	

\begin{document}


\title{Measurement of Single Muons at Forward Rapidity in $p+p$
Collisions at $\sqrt{s} = 200$\,GeV and Implications for Charm Production}

\newcommand{\abilene}{Abilene Christian University, Abilene, TX 79699, USA}
\newcommand{\acadsin}{Institute of Physics, Academia Sinica, Taipei 11529, Taiwan}
\newcommand{\banaras}{Department of Physics, Banaras Hindu University, Varanasi 221005, India}
\newcommand{\barc}{Bhabha Atomic Research Centre, Bombay 400 085, India}
\newcommand{\bnl}{Brookhaven National Laboratory, Upton, NY 11973-5000, USA}
\newcommand{\caucr}{University of California - Riverside, Riverside, CA 92521, USA}
\newcommand{\ciae}{China Institute of Atomic Energy (CIAE), Beijing, People's Republic of China}
\newcommand{\cns}{Center for Nuclear Study, Graduate School of Science, University of Tokyo, 7-3-1 Hongo, Bunkyo, Tokyo 113-0033, Japan}
\newcommand{\columbia}{Columbia University, New York, NY 10027 and Nevis Laboratories, Irvington, NY 10533, USA}
\newcommand{\dapnia}{Dapnia, CEA Saclay, F-91191, Gif-sur-Yvette, France}
\newcommand{\debrecen}{Debrecen University, H-4010 Debrecen, Egyetem t{\'e}r 1, Hungary}
\newcommand{\fsu}{Florida State University, Tallahassee, FL 32306, USA}
\newcommand{\gsu}{Georgia State University, Atlanta, GA 30303, USA}
\newcommand{\hiroshima}{Hiroshima University, Kagamiyama, Higashi-Hiroshima 739-8526, Japan}
\newcommand{\ihepprot}{Institute for High Energy Physics (IHEP), Protvino, Russia}
\newcommand{\isu}{Iowa State University, Ames, IA 50011, USA}
\newcommand{\jinrdubna}{Joint Institute for Nuclear Research, 141980 Dubna, Moscow Region, Russia}
\newcommand{\kaeri}{KAERI, Cyclotron Application Laboratory, Seoul, South Korea}
\newcommand{\kangnung}{Kangnung National University, Kangnung 210-702, South Korea}
\newcommand{\kek}{KEK, High Energy Accelerator Research Organization, Tsukuba-shi, Ibaraki-ken 305-0801, Japan}
\newcommand{\kfki}{KFKI Research Institute for Particle and Nuclear Physics (RMKI), H-1525 Budapest 114, POBox 49, Hungary}
\newcommand{\korea}{Korea University, Seoul, 136-701, Korea}
\newcommand{\kurchatov}{Russian Research Center ``Kurchatov Institute", Moscow, Russia}
\newcommand{\kyoto}{Kyoto University, Kyoto 606-8502, Japan}
\newcommand{\labllr}{Laboratoire Leprince-Ringuet, Ecole Polytechnique, CNRS-IN2P3, Route de Saclay, F-91128, Palaiseau, France}
\newcommand{\lawllnl}{Lawrence Livermore National Laboratory, Livermore, CA 94550, USA}
\newcommand{\losalamos}{Los Alamos National Laboratory, Los Alamos, NM 87545, USA}
\newcommand{\lpc}{LPC, Universit{\'e} Blaise Pascal, CNRS-IN2P3, Clermont-Fd, 63177 Aubiere Cedex, France}
\newcommand{\lund}{Department of Physics, Lund University, Box 118, SE-221 00 Lund, Sweden}
\newcommand{\muenster}{Institut f\"ur Kernphysik, University of Muenster, D-48149 Muenster, Germany}
\newcommand{\myongji}{Myongji University, Yongin, Kyonggido 449-728, Korea}
\newcommand{\nagasaki}{Nagasaki Institute of Applied Science, Nagasaki-shi, Nagasaki 851-0193, Japan}
\newcommand{\newmex}{University of New Mexico, Albuquerque, NM 87131, USA}
\newcommand{\nmsu}{New Mexico State University, Las Cruces, NM 88003, USA}
\newcommand{\ornl}{Oak Ridge National Laboratory, Oak Ridge, TN 37831, USA}
\newcommand{\orsay}{IPN-Orsay, Universite Paris Sud, CNRS-IN2P3, BP1, F-91406, Orsay, France}
\newcommand{\pnpi}{PNPI, Petersburg Nuclear Physics Institute, Gatchina, Russia}
\newcommand{\riken}{RIKEN (The Institute of Physical and Chemical Research), Wako, Saitama 351-0198, JAPAN}
\newcommand{\rikjrbrc}{RIKEN BNL Research Center, Brookhaven National Laboratory, Upton, NY 11973-5000, USA}
\newcommand{\saispbstu}{St. Petersburg State Technical University, St. Petersburg, Russia}
\newcommand{\saopaulo}{Universidade de S{\~a}o Paulo, Instituto de F\'{\i}sica, Caixa Postal 66318, S{\~a}o Paulo CEP05315-970, Brazil}
\newcommand{\seoulnat}{System Electronics Laboratory, Seoul National University, Seoul, South Korea}
\newcommand{\stonybrkc}{Chemistry Department, Stony Brook University, SUNY, Stony Brook, NY 11794-3400, USA}
\newcommand{\stonycrkp}{Department of Physics and Astronomy, Stony Brook University, SUNY, Stony Brook, NY 11794, USA}
\newcommand{\subatech}{SUBATECH (Ecole des Mines de Nantes, CNRS-IN2P3, Universit{\'e} de Nantes) BP 20722 - 44307, Nantes, France}
\newcommand{\tenn}{University of Tennessee, Knoxville, TN 37996, USA}
\newcommand{\titech}{Department of Physics, Tokyo Institute of Technology, Tokyo, 152-8551, Japan}
\newcommand{\tsukuba}{Institute of Physics, University of Tsukuba, Tsukuba, Ibaraki 305, Japan}
\newcommand{\vandy}{Vanderbilt University, Nashville, TN 37235, USA}
\newcommand{\waseda}{Waseda University, Advanced Research Institute for Science and Engineering, 17 Kikui-cho, Shinjuku-ku, Tokyo 162-0044, Japan}
\newcommand{\weizmann}{Weizmann Institute, Rehovot 76100, Israel}
\newcommand{\yonsei}{Yonsei University, IPAP, Seoul 120-749, Korea}
\affiliation{\abilene}
\affiliation{\acadsin}
\affiliation{\banaras}
\affiliation{\barc}
\affiliation{\bnl}
\affiliation{\caucr}
\affiliation{\ciae}
\affiliation{\cns}
\affiliation{\columbia}
\affiliation{\dapnia}
\affiliation{\debrecen}
\affiliation{\fsu}
\affiliation{\gsu}
\affiliation{\hiroshima}
\affiliation{\ihepprot}
\affiliation{\isu}
\affiliation{\jinrdubna}
\affiliation{\kaeri}
\affiliation{\kangnung}
\affiliation{\kek}
\affiliation{\kfki}
\affiliation{\korea}
\affiliation{\kurchatov}
\affiliation{\kyoto}
\affiliation{\labllr}
\affiliation{\lawllnl}
\affiliation{\losalamos}
\affiliation{\lpc}
\affiliation{\lund}
\affiliation{\muenster}
\affiliation{\myongji}
\affiliation{\nagasaki}
\affiliation{\newmex}
\affiliation{\nmsu}
\affiliation{\ornl}
\affiliation{\orsay}
\affiliation{\pnpi}
\affiliation{\riken}
\affiliation{\rikjrbrc}
\affiliation{\saispbstu}
\affiliation{\saopaulo}
\affiliation{\seoulnat}
\affiliation{\stonybrkc}
\affiliation{\stonycrkp}
\affiliation{\subatech}
\affiliation{\tenn}
\affiliation{\titech}
\affiliation{\tsukuba}
\affiliation{\vandy}
\affiliation{\waseda}
\affiliation{\weizmann}
\affiliation{\yonsei}
\author{S.S.~Adler}	\affiliation{\bnl}
\author{S.~Afanasiev}	\affiliation{\jinrdubna}
\author{C.~Aidala}	\affiliation{\bnl}
\author{N.N.~Ajitanand}	\affiliation{\stonybrkc}
\author{Y.~Akiba}	\affiliation{\kek} \affiliation{\riken}
\author{J.~Alexander}	\affiliation{\stonybrkc}
\author{R.~Amirikas}	\affiliation{\fsu}
\author{L.~Aphecetche}	\affiliation{\subatech}
\author{S.H.~Aronson}	\affiliation{\bnl}
\author{R.~Averbeck}	\affiliation{\stonycrkp}
% ...
\collaboration{PHENIX Collaboration} \noaffiliation

\date{\today}

\begin{abstract}

Muon production at forward rapidity ($1.5 \le |\eta| \le 1.8$) has
been measured by the PHENIX experiment over the transverse momentum
range $1 \le p_T \le 3$\,GeV/c in $\sqrt{s} = 200$\,GeV $p+p$
collisions at the Relativistic Heavy Ion Collider. After statistically
subtracting contributions from light hadron decays an excess remains which
is attributed to the semileptonic decays of hadrons carrying heavy
flavor, {\it i.e.}  charm quarks or, at high $p_T$, bottom quarks. The
resulting muon spectrum from heavy flavor decays is compared to PYTHIA
and a next-to-leading order perturbative QCD calculation. PYTHIA is
used to determine the charm quark spectrum that would produce the
observed muon excess. The corresponding differential cross
section for charm quark production at forward rapidity is determined
to be $d\sigma_{c\bar{c}}/dy|_{y=1.6} = 0.243 \pm 0.013 {\rm (stat.)}
\pm 0.105 {\rm (data~syst.)}~^{+0.049}_{-0.087}{\rm
(PYTHIA~syst.)}$\,mb.

\end{abstract}


\maketitle

\section{Introduction}
\label{sec:intro}

Measurements of heavy quark production in proton-proton ($p+p$)
interactions at collider energies serve as important tests for
perturbative Quantum ChromoDynamics (pQCD). Bottom production at the
Tevatron collider ($\sqrt{s}=1.8$ and 1.96\,TeV/c)~\cite{abbott:2000,
CDFBottom} is reasonably well described by a recent Fixed Order
Next-to-Leading Logarithm (FONLL) calculation~\cite{cacciari:2004,
cacciari:1998, mangano:2004}. Charm production at FNAL, which has only
been measured at relatively high $p_T$ ($> 5$\,GeV/c), is $\approx$
50\% higher than the FONLL prediction~\cite{CDFCharm}. However,
theoretical and experimental uncertainties are large, such that
significant disagreement between theory and data cannot be
claimed. 

Measurements at Brookhaven National Laboratory's Relativistic Heavy
Ion Collider (RHIC), by both the PHENIX and STAR experiments, have
provided a wealth of information on mid-rapidity open charm production
in collisions at $\sqrt{s_{NN}} = 130$\,GeV ($p+p$) and $\sqrt{s_{NN}}
= 200$\,GeV ($p+p$, $d+Au$, and $Au+Au$) down to $p_T \approx
0.5$\,GeV/c. Semileptonic decay of produced charm quarks is the primary
source of high $p_T$ leptons after contributions from known (light
hadron) sources are subtracted. Both
PHENIX~\cite{ppg011,ppg035,ppg037,kelly:2004,ppg056,butsyk:2005,ppg065,ppg040,sakai:2005,
ppg066} and STAR~\cite{adams:2005,abelev:2006} have made statistical
measurements of charm production via single-electron spectra. STAR has
also made a direct measurement of charm production through
reconstruction of hadronic decay modes of $D$
mesons~\cite{adams:2005}. In $p+p$ collisions at $\sqrt{s_{NN}} =
200$\,GeV PHENIX finds $d\sigma_{c\bar{c}}/dy|_{y=0} = 0.123 \pm 0.012
\rm{(stat.)} \pm 0.045 \rm{(syst.)}$\,mb~\cite{ppg065}. STAR finds a somewhat higher
central value, $d\sigma_{c\bar{c}}/dy|_{y=0} = 0.30 \pm 0.04
\rm{(stat.)} \pm 0.09 \rm{(syst.)}$\,mb~\cite{adams:2005}, but the two measurements are
consistent within the stated errors. Both measurements are noticeably
(2-4$\times$) higher than PYTHIA (a leading order pQCD event
generator)~\cite[see experimental references for specific parameter
sets]{Sjostrand:2001} and FONLL~\cite{cacciari:2005}. Again,
quantitative disagreement cannot be established with current
experimental and theoretical errors. However, we note that there is
some debate about whether charm quarks are heavy enough to be reliably
treated by pQCD~\cite{Mangano:1993}.

Such measurements also serve as an important baseline for charm
production in proton-nucleus or deuteron-nucleus ($p+A$ or $d+A$), and
nucleus-nucleus ($A+B$) collisions~\cite{vogt:2002, vogt:2003,
cassing:2001, bratkovskaya:2003}. In the absence of any nuclear
effects, charm production (since it is a point-like process) is
expected to scale with the number of binary nucleon-nucleon collisions
($N_{coll}$), which depends on the impact parameter of the nuclear
collision and can be obtained from a Glauber
calculation~\cite{Wong:Textbook}. The degree of scaling for any given
centrality bin is quantified by the nuclear modification factor:
\begin{equation}
R_{AB} = \frac{1}{N_{coll}^{AB}} \times \frac{dN^{AB}/dy}{dN^{pp}/dy}.
\end{equation}
Deviations from this scaling ($R_{AB} \neq 1$) in $p+A$ or $d+A$
collisions quantify cold nuclear matter effects (such as initial state
energy
loss~\cite{Johnson:2001,Bodwin:1981,Bodwin:1985,Bodwin:1989,Kopeliovich:1984,Gavin:1992},
and shadowing~\cite{Guzey:2004, dokshitzer:2001, Mclerran:1994a,
Mclerran:1994b, Ashman:1988}). Any such deviation must be understood
so that in $A+B$ collisions contributions to $R_{AB} \neq 1$ from hot
nuclear matter effects (such as in-medium energy loss~\cite[and
references therein]{baier:2000}) and cold nuclear matter effects can
be disentangled. In $d+Au$ collisions both PHENIX and STAR find little
or no effect of cold nuclear matter on charm production ($R_{dAu}
\approx 1$ over the measured lepton
$p_T$~\cite{kelly:2004,adams:2005}). This contrasts with measurements
of open charm in $Au+Au$ collisions: although the {\em total} charm
production appears to scale with $N_{coll}$~\cite{ppg035}, there is a
strong suppression of lepton spectra for $p_T > 2$\,GeV/c that
increases with
centrality~\cite{ppg056,butsyk:2005,abelev:2006}. Furthermore the
elliptical flow of non-photonic single electrons, as measured by
PHENIX in $Au+Au$ collisions~\cite{ppg040,sakai:2005,ppg066}, implies that the
charm quarks interact strongly with the created medium.

Finally, since the initial formation of open and closed charm are both
sensitive to initial gluon densities~\cite{appel:1992, muller:1992},
open charm production serves as an appropriate normalization for
$J/\psi$ production. The production of $J/\psi$ mesons is expected to
be sensitive to the production of a quark gluon plasma (QGP), should
one be formed in $A+B$ collisions~\cite{matsui:1986, satz:2005,
thews:2005, grandchamp:2004, andronic:2003, kostyuk:2003}. In order to
understand $J/\psi$ production differences in $A+B$ collisions
compared to $p+p$ and $p+A$ collisions it is important to take into
account any differences in the charm quark production in each of the
different systems.

Until now, open charm measurements at RHIC have been limited to
mid-rapidity. Measurements at forward rapidity are interesting for a
variety of reasons. First is the need to constrain theoretical
calculations over a wide kinematic range. The importance of this is
demonstrated by the D0 measurement of bottom production at large
rapidity ($\sqrt{s} = 1.8$ TeV, $p_T > 5$\,GeV/c, $2.4 < y_{\mu} <
3.2$), as deduced from the production of high $p_T$
muons~\cite{abbott:2000}. Significant theoretical improvements
resulted from the effort to reduce what was, initially, a discrepancy
between theory and experiment that increased with increasing
rapidity~\cite{mangano:2004}. Second, significant cold nuclear effects
have been seen in RHIC collisions at forward
rapidity. PHENIX~\cite{ppg036}, BRAHMS~\cite{Arsene:2005,
Arsene:2004}, and STAR~\cite{starfpi0} have all measured light hadron
production in $d+Au$ collisions at forward rapidity and have found
significant deviations from $R_{dAu} = 1$. It will be interesting to
see whether charm production follows a similar pattern. Finally, open
charm production at forward rapidity needs to be understood to fully
interpret PHENIX $J/\psi$ measurements at forward
rapidity~\cite{ppg017,ppg038,pereira:2005,cassing:2001,bratkovskaya:2003}.

In this paper we report on the measurement of muon production at
forward rapidity ($1.5 \le |\eta| \le 1.8$), in the range $1 < p_T <
3$\,GeV/c, in $\sqrt{s} = 200$\,GeV $p+p$ collisions by the PHENIX
experiment. The upper limit of the $p_T$ range is determined by
available statistics. The vertex-independent muon yield is
statistically extracted by calculating and subtracting contributions
from light mesons ($\pi$'s and $K$'s) which decay into a muon, and
hadrons which penetrate through the muon arm absorber material. In the
absence of new physics, and in the $p_T$ range measured in this
analysis, such muons come dominantly from the decay of hadrons
containing a charm quark (with small contributions from decays of
hadrons containing a bottom quark and decays of light-vector
mesons). PYTHIA is used to determine the charm quark spectrum that
would produce the observed vertex-independent muon spectrum, and from
this we obtain the differential cross section of charm quark
production at forward rapidity.

\begin{figure*}[thb]
\includegraphics[width=0.80\linewidth]{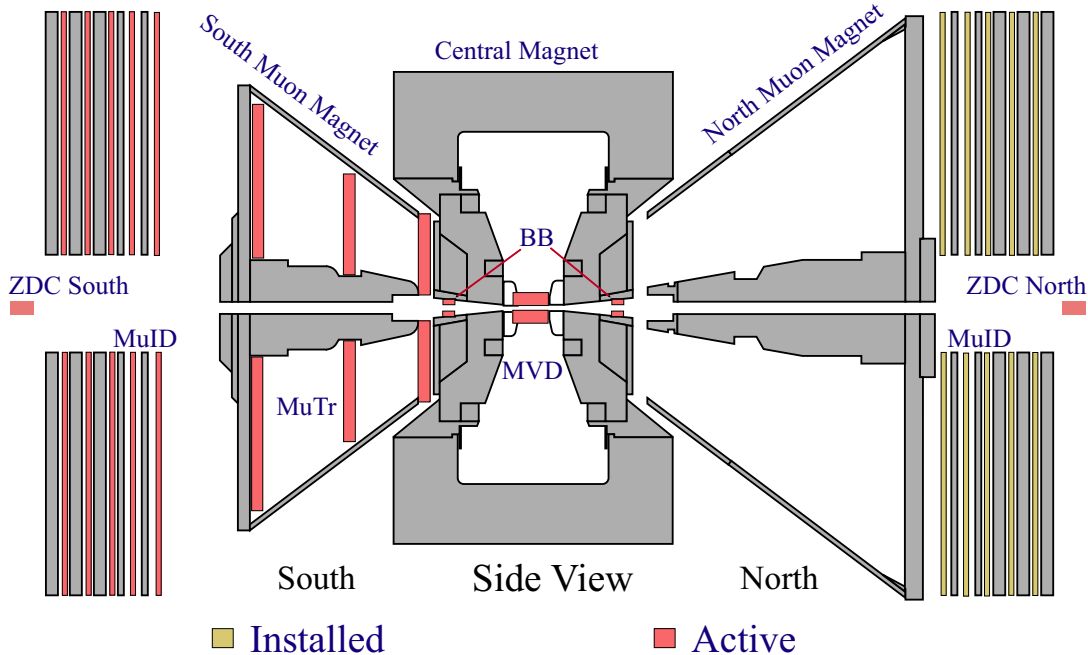}
\caption{(Color online) PHENIX experimental layout during the 2001/2
run period.}
\label{fig:expfig}
\end{figure*}

The remainder of this paper is organized as follows: In
Section~\ref{sec:experiment} we describe the PHENIX experimental
apparatus, with special emphasis on the muon arm detectors. In
Section~\ref{sec:method} we describe the methodology used to extract
the vertex-independent muon signal. This section includes details on
the run, event and track selection criteria; values obtained for
contributions to the muon yield from abundant light hadrons, which are
subtracted to obtain the vertex-independent muon yield; and details on
the systematic error analysis. In Section~\ref{sec:xsec} we extract
the differential cross section for charm production at $y=1.6$,
integrated over $p_T$. Finally, in Section~\ref{sec:conclusion} we
compare to other measurements, draw conclusions, and discuss the
prospects for such measurements with improved data sets currently
under analysis.

\section{The PHENIX Experiment}
\label{sec:experiment} 

The PHENIX experiment~\cite{phenix_nim}, shown in
Figure~\ref{fig:expfig}, is a large multipurpose set of detectors
optimized for measuring relatively rare electromagnetic probes
(photons, muons, and electrons) of the spin structure of the proton
and of the hot dense matter created in ultrarelativistic heavy ion
collisions. The data acquisition system and multilevel triggers are
designed to handle the very different challenges presented by $p+p$
collisions (relatively small events at very high rates) and $Au+Au$
collisions (very large events at relatively low rates) with little or
no deadtime~\cite{phenix_daq, phenix_online}. Event characterization
devices, such as the Beam-Beam Counters~\cite{phenix_inner} used in
this analysis, provide information on the vertex position, start time,
and centrality of the collision. The two muon arms cover $1.2 < |\eta|
< 2.4$ in pseudorapidity and $\delta\phi = 2\pi$ in azimuth. The two
central arms, which each cover $|\eta| < 0.35$ and $\delta\phi =
\pi/2$, are not used in this analysis.

The Beam-Beam Counters (BBCs)~\cite{phenix_inner} each consist of 64
quartz radiators instrumented with mesh dynode PMTs and arranged in a
cylinder coaxial with the beam. The BBCs are placed on either side of
the collision vertex and cover $3.0 < |\eta| < 3.9$. Each channel has
a dynamic range extending to 30\,MIPs. The BBCs measure the arrival
times of particles on both sides of the collision vertex, $t_S$ and
$t_N$. From the average of these times we determine the event start
time. From their difference we obtain the position of the vertex along
the beam direction, $z_{vtx}$. The BBCs also provide the minimum bias
interaction trigger, which requires that there be at least one hit in
each BBC and that $|z_{vtx}| < 38$\,cm.

The muon arms~\cite{phenix_muon} are coaxial with the beam on opposite
sides of the collision vertex. By convention the arm on the South
(North) end of the interaction region is assigned negative (positive)
$z$ coordinates and rapidity. For the 2001/2 run period, in which the
data for this paper were collected, only the South muon arm was
operational. Each muon arm is comprised of a Muon Tracker (MuTR) and a
Muon Identifier (MuID). The MuTR makes an accurate measurement of
particle momenta. The MuID allows coarse resolution track
reconstruction through a significant amount of steel
absorber. Together the muon arm detectors provide significant pion
rejection ($>250:1$, increasing with decreasing momentum) through a
momentum/penetration-depth match.

Before ever reaching the MuTR detectors a particle must pass through
the pre-MuTR absorber: 20\,cm of copper (the nosecone) plus 60\,cm of
iron (part of the MuTR magnet). The nominal nuclear interaction
lengths of iron and copper are $\lambda_I^{Fe} = 16.7$\,cm and
$\lambda_I^{Cu} = 15.3$\,cm (although this varies with particle
species and energy, see Section~\ref{sec:punch}). Therefore the
pre-MuTR absorber presents a total thickness of
$4.9\lambda_I/\cos\theta$, where $\theta$ is the polar angle of a
particle's trajectory. This absorber greatly reduces the MuTR
occupancy and provides the first level of pion rejection.

Each MuTR arm consists of three stations of cathode strip chambers
installed in an eight-sided conical magnet~\cite{phenix_magnet}. The
radial magnetic field (${\bf \int B \cdot dl} = 0.72$\,T$\cdot$m at 15
degrees, $B(\theta) \approx B(15^\circ)\tan(\theta)/\tan(15^\circ)$)
bends particles in the azimuthal direction. Each station occupies a
plane perpendicular to the beam axis and consists of multiple
ionization gaps (3 gaps for the two stations closest to the collision
vertex, 2 gaps for the last station) which have their charge imaged on
two cathode strip planes oriented with a small stereo angle to provide
two-dimensional information. An ionizing particle typically fires
three adjacent strips in each orientation. A fit to the charge
distribution on adjacent strips provides a position measurement with a
resolution of $\sigma \approx 100\,\mu$m in the bend direction.  The
MuTR achieves a momentum resolution of $\sigma_p/p \approx 5\%$ over
the analyzed kinematic range. The resolution is approximately
independent of momentum due to the significant contribution from
energy loss fluctuations in the pre-MuTR absorber, which falls as
$1/p$, and which counters the more familiar linear momentum dependence seen
for particles tracked through a ``thin'' detector.

Each MuID arm consists of five steel absorber plates interleaved with
Iarocci tubes (operated in proportional mode) and specialized shielding
to reduce backgrounds not originating from the collision vertex. Gaps
are labeled 0--4 proceeding downstream from the collision point.

The Iarocci tubes, which are between 2.5 and 5\,m in length, have
eight 1\,cm$^2$ square cells, each consisting of a three-sided ground
electrode and an anode wire, mounted inside a PVC gas enclosure. A
readout channel (``two-pack'') is formed by wire-ORing the 16 anode
wires of two tubes which are mounted in planes perpendicular to the
beam axis and staggered by half of a cell width (0.5\, cm). This
provides redundancy, eliminates geometric inefficiency due to the cell
walls, and reduces the maximum drift time for charge
collection. Digital readout of the two-pack signals provides a coarse
one-dimensional hit position ($\sigma = 9\,{\rm
cm}/\sqrt{12} = 2.6\,{\rm cm}$). The tubes in each gap are mounted in
six individual panels, each of which contains two layers of two-packs
(horizontally and vertically oriented), thus providing two-dimensional
information.

\begin{figure*}[thb]
\includegraphics[width=0.9\linewidth]{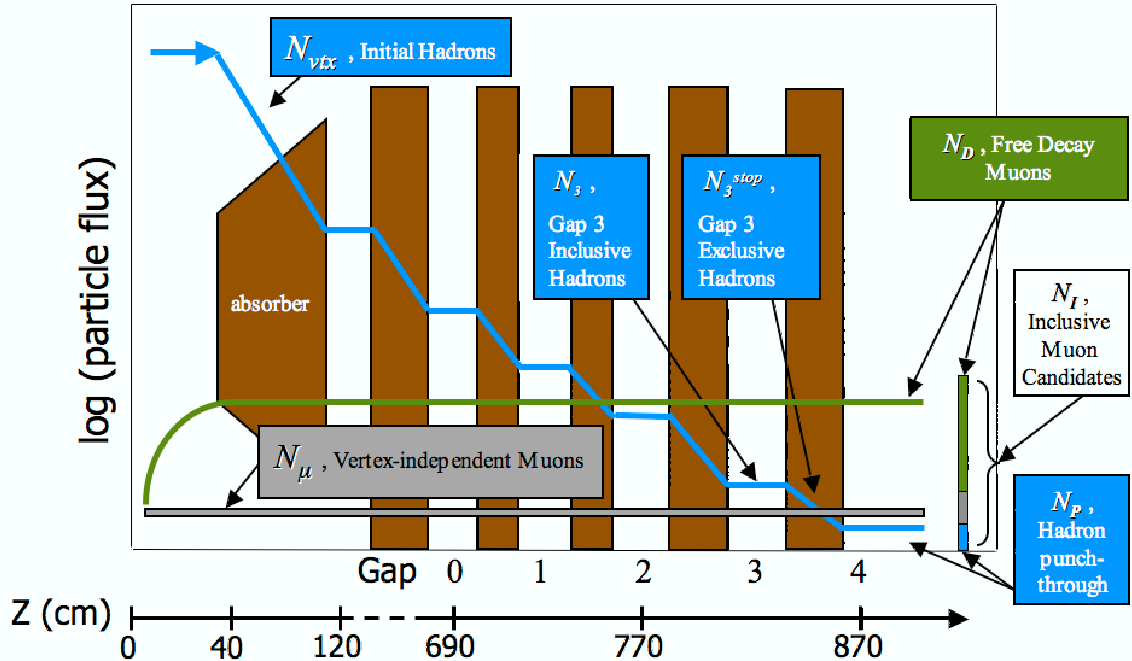}
\caption{(Color online) Schematic depiction of the relative flux of
different components of the inclusive muon candidate yield as a
function of flightpath into the muon arm absorber (the event vertex is
at $z_{vtx}=0$). See text for details.}
\label{fig:penetration}
\end{figure*}

The first MuID absorber plate (thickness = 20\,cm - South; 30\,cm -
North) also serves as the return yoke of the MuTR magnet. Successive
plates (identical for the two arms) are 10, 10, 20 and 20\,cm thick,
thus totaling $4.8\lambda_I/\cos\theta$
($5.4\lambda_I/\cos\theta$) for the South (North) arm. Due to
ionization energy loss a particle must have a momentum at the vertex
which exceeds $2.31\cos\theta$\,GeV/c ($2.45\cos\theta$\,GeV/c) to
penetrate to the most downstream MuID gap of the South (North) arm.

Steel plates surrounding the beam pipe guard against backgrounds
caused by low-angle beam-beam collision products which scrape the beam
pipe near the MuID $z$-location (7-9\,m) or shine off the RHIC DX
magnets immediately downstream of each MuID arm. Steel
blocks in the RHIC tunnels guard against penetrating radiation
generated by the incoming beams scraping against beamline components
(primarily the final focusing quadrupoles).

The MuID also contributes information to the first-level trigger
decision. For the 2001/2 run, during which the data for this analysis
were collected, a relatively coarse trigger was implemented using
LeCroy 2372 Memory Lookup Units (MLUs). Each gap was divided into
quadrants with a horizontal and vertical split going through the beam
axis. Signals from tubes in an individual gap/orientation (layer) and
quadrant were logically combined. Only gaps 0,2,3 and 4 were used in
the trigger due to the 16-bit input limitation of the MLUs. The
penetration depth required for the trigger to fire was
programmable. The {\em MuID-1Deep} trigger fired if more
than 6 out of 8 layers in a particular quadrant were hit (indicating
the possibility that the event contained a particle penetrating to
MuID gap 4). The {\em MuID-1Shallow} trigger fired if 3 of
the 4 most shallow layers (horizontal and vertical layers in gaps 0
and 2) were hit for a particular quadrant.

\section{Method for Extraction of Muons from Charm Decay}
\label{sec:method}

Inclusive muon candidates, $N_I$, are those particles which are
successfully reconstructed to the last MuID gap (gap 4). These consist
of four components: 1) ``free-decay muons'', $N_D$, which result from
the decay of light hadrons ($\pi$ and $K$ mesons) before reaching the
pre-MuTR absorber, 2) ``punchthrough hadrons'', $N_P$, which penetrate
the entire detector and are thus misidentified as muons 3)
``background tracks'', $N_B$, which in $p+p$ collisions are dominated
by hadrons which decay into a muon after reaching the MuTR, and 4)
``vertex-independent muons'', $N_{\mu}$, which are primarily due to
the decay of heavy flavor mesons.

Figure~\ref{fig:penetration} shows a schematic depiction of the
relative yield per event of these different contributions as a
function of flightpath into the muon arms, as described below. 

The number of hadrons is large and essentially independent of
flightpath until the first absorber layer is reached. In each absorber
layer these hadrons undergo strong interactions with a probability
$1-\exp(-L/\lambda)$, where $L$ is the length of absorber material
traversed, and $\lambda$ is the species and $p_T$-dependent nuclear
interaction length determined in Section~\ref{sec:punch}. Most of these
interacting hadrons are effectively eliminated as possible muon
candidates. However, a small fraction of hadrons penetrate the entire
absorber thickness. These punchthrough
hadrons are indistinguishable from muons.

The decay lengths for $\pi$'s (c$\tau = 780$\,cm) and $K$'s (c$\tau =
371$\,cm) are long compared to the flightpath from the vertex to the
absorber. Therefore, the fraction of decay muons from these sources is
relatively small, but increases linearly with the flightpath until the
first absorber layer is reached. A hadron which decays prior to the
pre-MuTR absorber into a muon that is inside the detector acceptance
is indistinguishable from a muon originating at the vertex. After the
first absorber layer the number of free-decay muons remains constant
by definition.

Hadrons which decay in the MuTR volume are a relatively small
contribution since most are absorbed prior to reaching the MuTR, the
Lorentz-dilated decay lengths are long compared to the length of the
MuTR volume (South $\approx 280$\,cm, North $\approx 420$\,cm), and a
particle which decays in the MuTR is less likely to be
reconstructed. Such tracks are partially accounted for in the
calculation of punchthrough hadrons (see Section~\ref{sec:punch}) and
the remaining fraction falls under the category of background tracks
(Section~\ref{sec:back}). This small contribution is not shown.

Without a high-resolution vertex detector muons from various sources
(the decay of open heavy flavor hadrons, the decay of quarkonia, the
decay of light vector mesons, and Drell-Yan production) originate
indistinguishably close to the collision vertex. Thus their yield is
independent of the flightpath and independent of the vertex position.
Since inclusive muon candidates, by definition, penetrate to MuID gap
4, we measure the combined yield at $z \approx 870$\,cm.

\begin{figure}[hbt]
\includegraphics[width=1.0\linewidth]{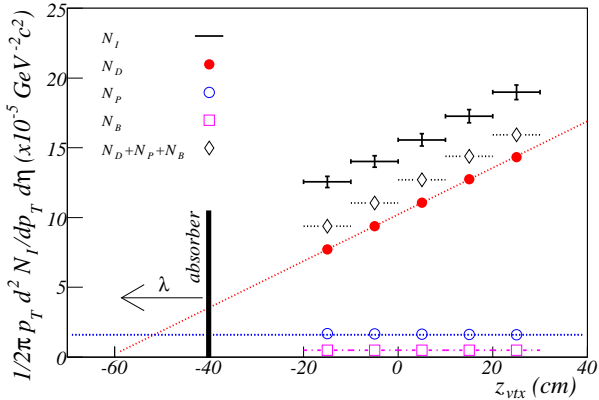}
\caption{Sample $z_{vtx}$ distribution of different components of the
  inclusive muon candidate yield (measured data for both charge signs
  over the range $1.0 < p_T < 1.2$\,GeV/c). Crosses show inclusive
  muon candidates, filled circles show free-decay muons, open circles
  show punchthrough hadrons, open squares show background tracks, and
  open diamonds show the sum of these three hadronic sources. The
  vertex-independent muon yield is obtained from the difference
  between the yield of inclusive muon candidates and the yield of
  light hadronic sources. }
\label{fig:composition}
\end{figure}

Figure~\ref{fig:composition} shows a sample distribution of the
inclusive muon candidate yield as a function of collision vertex
($z_{vtx}$), and its decomposition into the four different
contributions. The yield of free-decay muons is seen to have a linear
dependence that is set to 0 at $z_{vtx} = z_{abs}-\lambda_D$. Here
$z_{abs}=-40$\,cm is the upstream face of the pre-MuTR absorber
(indicated by the thick solid line), and $\lambda_D$ is the effective
absorption length, beyond which there are no free-decay
muons. $\lambda_D$ was found to be nearly identical to the species and
$p_T$-dependent nuclear interaction lengths determined in
Section~\ref{sec:punch}. Muons originating from meson decays
downstream of this location have no $z_{vtx}$ dependence. The fraction
not accounted for in the calculation of the punchthrough hadron yield
forms the small contribution from background tracks. The yield of
punchthrough hadrons and vertex-independent muons also have no
$z_{vtx}$ dependence. Note that the ratio of different contributions
to the inclusive muon candidate spectrum is $p_T$ dependent.

In order to extract the cross section for charm production we first
need to determine the yield of vertex-independent muons,
$N_{\mu}(p_T)$, the amount beyond that due to light hadrons and fake
backgrounds. As described in Section~\ref{sec:datareduction}, we
select good runs, events, and tracks, and restrict our acceptance to
regions where the detector performance was optimal, and the acceptance
vs.~$z_{vtx}$ was nearly constant. Next, as described in
Sections~\ref{sec:acceff} and~\ref{sec:inclus}, we obtain the yield of
inclusive muon candidates vs.~$p_T$ and $z_{vtx}$, corrected for
acceptance and efficiency: $N_I(p_T, z_{vtx})$. In
Section~\ref{sec:cocktail} we describe a data-driven hadron generator.
This generator is used in Section~\ref{sec:decay}, in which we
describe how the vertex dependence of the inclusive muon candidate
yield allows us to obtain the yield of muons from light-meson decay before
the pre-MuTR absorber, similarly corrected and binned: $N_D(p_T,
z_{vtx})$. This generator is also used in Section~\ref{sec:punch}, in
which we describe how we use hadrons which stop in MuID gap 3 (the
penultimate gap), together with simulations of hadron penetration in
the MuID absorber, to obtain the yield of punchthrough hadrons in MuID
gap 4: $N_P(p_T, z_{vtx})$. The yield of fake tracks, $N_B(p_T,
z_{vtx})$, determined from simulations described in
Section~\ref{sec:back}, is found to be small.

The yield of vertex-independent muons is determined by subtracting the
contributions from light hadrons and fake backgrounds and averaging over
$z_{vtx}$ bins:
\begin{widetext}
\begin{equation}
\label{eq:muonsources}
N_{\mu}(p_T) = \frac{1}{N_{z_{vtx}}}\sum_{j=1}^{N_{z_{vtx}}} N_I(p_T,
z^j_{vtx}) - N_D(p_T, z^j_{vtx}) - N_P(p_T, z^j_{vtx}) - N_B(p_T,
z^j_{vtx}),
\end{equation}
\end{widetext}
where $d^2/2 \pi p_T dp_T d\eta$ is implicit in all terms of the equation.

We convert this into a cross section via
\begin{equation}
\label{eq:ptspectrum}
\frac{d^2\sigma_{\mu}(p_T)}{2 \pi p_T dp_T dy} =
\frac{\sigma_{BBC}^{pp}}{\varepsilon_{BBC}^{c,\bar{c}\rightarrow\mu}} 
\frac{d^2N_{\mu}(p_T)}{2 \pi p_T dp_T d\eta}.
\end{equation}
Here $\sigma_{BBC}^{pp}$ is the cross section of the BBC trigger for $p+p$
interactions and $\varepsilon_{BBC}^{c,\bar{c}\rightarrow\mu}$ is the
efficiency of the BBC trigger for events in which a charm quark is
created and decays into a muon. Substituting $\eta \rightarrow y$
introduces negligible error due to the small mass of the muon, the
only component left after the subtraction. As described in
Section~\ref{sec:syst}, systematic errors are determined for each
component and combined into a term that applies to the overall
normalization and a term that applies to the $p_T$ dependence of the
spectrum.

We use PYTHIA to derive the $p_T$-dependent differential cross section
for the production of charm quarks responsible for the
vertex-independent muon yield. This procedure is very similar to that
in references~\cite{ppg011,ppg035,ppg037,kelly:2004,ppg056,butsyk:2005,ppg065}, and is described in
detail, along with the associated systematic error analysis, in
Section~\ref{sec:xsec}.

\subsection{Data Reduction}
\label{sec:datareduction}

\subsubsection{Data Sets and Triggering}
\label{sec:runsel}

Runs were selected for this analysis based on stable detector
operation using the same criteria as an earlier analysis of $J/\psi$
production~\cite{ppg017}. Further runs were eliminated due to the
presence of large beam-related backgrounds entering the {\em back} of
the detector.

We select only those events in the vertex range $-20 < z_{vtx} <
30$\,cm. This minimizes the $z_{vtx}$ dependence of the detector
acceptance and allows us to treat the amount of absorber material as a
simple function of polar angle, ignoring complications in the pre-MuTR
absorber near the beampipe.

The decision to collect an event was made by the Local Level-1
Trigger~\cite{phenix_daq} within 4\,$\mu$s of the collision. Input to
the trigger decision was given by the BBC (collision with a valid
event vertex) and the MuID (reconstructed penetrating track). Each
trigger could be independently scaled down, so that it occupied a
predetermined fraction of the available bandwidth, by selecting every
$N_i^{th}$ instance, where $N_i$ is the scaledown factor for the
$i^{th}$ trigger. Three different data sets were selected from the
good runs for different aspects of the data analysis:

\begin{itemize}

\item {\em BBC}: To extract $N_D$ we need to measure the $z_{vtx}$
dependence of $N_I$. For this we need the unbiased collision vertex
distribution, which we obtain from a set of events collected with the
{\em BBC} trigger: $N_{BBCN} > 1~\&\&~N_{BBCS} > 1~\&\&~|z_{vtx}| <
38$\,cm, where $N_{BBCN}$ and $N_{BBCS}$ are the number of hits in the
North and South BBC respectively. $\sigma^{pp}_{BBC}$ was found to be
$21.8 \pm 2.1$\,mb using a van der Meer scan~\cite{ppg024}. There were
$1.72 \times 10^7$ {\em BBC} triggered events passing our vertex
selection criteria in this data set, corresponding to a sampled
luminosity of $\int L\,dt = 0.79$\,nb$^{-1}$.

\item{\em MuID-1Deep}~$\&\&$~{\em BBC} ({\em M1D}): In order to
extract $N_I$, $N_D$ and $N_B$ we used events selected with the {\em
M1D} trigger, which enriched the sample of events with tracks
penetrating to MuID gap 4. For the {\em M1D} and {\em BBC} data sets we
used identical run selection criteria. The total number of sampled
{\em BBC} triggers for this data set was $5.77 \times 10^8$,
corresponding to a sampled luminosity of $\int L\,dt =
26.5$\,nb$^{-1}$.

\item{\em MuID-1Shallow}~$\&\&$~{\em BBC} ({\em M1S}): In order to
extract $N_P$ we need a data set which provides an unbiased
measurement of the number of particles which penetrate to MuID gap
3. Since the {\em M1D} trigger required tracks to penetrate to MuID
gap 4 it could not be used. Instead we used the {\em M1S} trigger,
which only used information from MuID gaps 0-2. We used a subset of
runs for which the scaledown factor for this trigger was only 10,
corresponding to a sampled luminosity of $\int L\,dt =
1.72$\,nb$^{-1}$.

\end{itemize}

\subsubsection{Track Selection}
\label{sec:tracksel}

\begin{figure}[thb]
\includegraphics[width=1.0\linewidth]{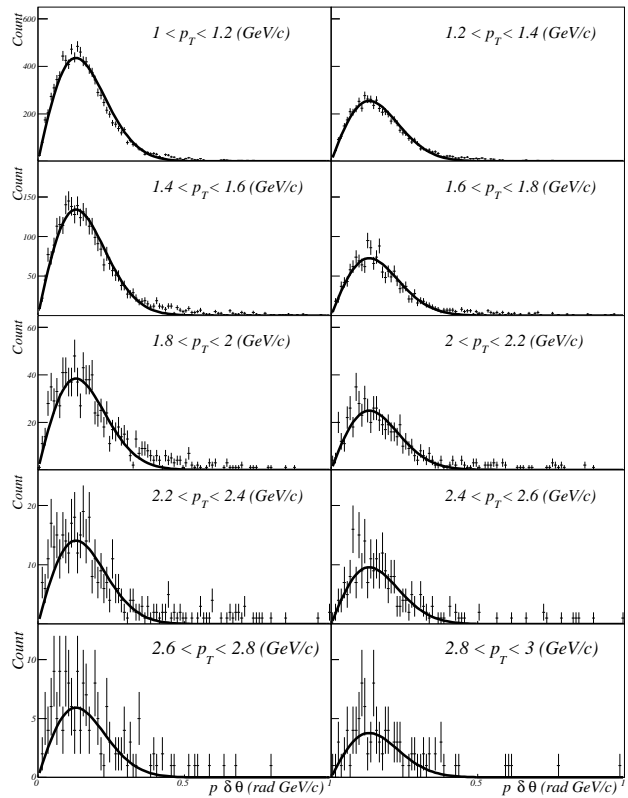}
\caption{The scaled angular deflection is the difference in a
  particle's polar angle caused by passage through the pre-MuTR
  absorber scaled by the particle's momentum, $p\delta\theta$. For
  muons (and hadrons not undergoing a strong interaction in the
  pre-MuTR absorber) one expects the distribution of this quantity to
  be well described by the standard multiple scattering formula. The
  different panels show $p\delta\theta$ for different $p_T$ bins with
  fits (normalization only) to the expected distribution.}
\label{fig:scaleddeflection}
\end{figure}

\begin{table}[tbh]
\caption{Road and track cuts. Here $D_p$ is the penetration depth,
defined to be the most downstream MuID gap with at least one hit (from
the horizontal or vertical layer) associated with the track; $z$ is
the coordinate along the beam; $x$ and $y$ are transverse to each
other and to the beam axis; the vertex cut refers to the transverse
position of the MuID road projected to the $xy$ plane at $z=0$; and
the slope cut refers to the direction cosine of the road in each
transverse direction.}
\label{tab:roadcuts}
\begin{ruledtabular}\begin{tabular}{ll}
           & \# Associated MuID hits, $N_{\it MuID} > 6$ \\
           & \hspace*{0.15 in} (out of a possible $2D_p$) \\ \hline
Road cuts  & Vertex cut, $\sqrt{x^2 + y^2} < 100$\,cm @ $z = 0$ \\
           & Slope cut, $\sqrt{(\frac{dx}{dz})^2 + (\frac{dy}{dz})^2} > 0.25$ \\
           & $\ge 1$ associated hit in MuID gap 4 \\ \hline
Track cuts & Track fit quality, $\chi^2/{\rm dof} \leq 10$ \\
           & \# Associated MuTR hits, $N_{\it MuTR} > 12$ \\
           & \hspace*{0.15 in} (out of a possible 16) \\
\end{tabular}\end{ruledtabular}
\end{table} 

\begin{figure}[tbh]
\includegraphics[width=1.0\linewidth]{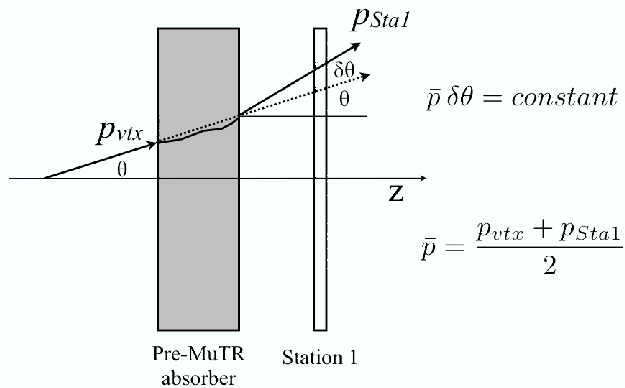}
\caption{The angular deflection, $\delta\theta$ is the angular
difference between the reconstructed particle trajectory at the
collision vertex and at MuTR station 1. The momentum used to scale
$\delta\theta$ is the average of the momentum reconstructed inside the
MuTR magnet ($p_{sta 1}$) and the momentum extrapolated to the vertex
($p_{vtx}$).}
\label{fig:dtheta}
\end{figure}

The Muon arm reconstruction algorithm starts by finding ``roads''
(collections of hits in the MuID which form straight, two-dimensional
lines) and then combining them with hits in the MuTR to form
``tracks''. We apply strict cuts on both road and track parameters in
order to reduce backgrounds, see Table~\ref{tab:roadcuts}.

The resulting purity of the selected tracks is demonstrated in
Figure~\ref{fig:scaleddeflection}. This figure shows $p\delta\theta$,
the angular deviation through the pre-MuTR absorber, scaled by the
particle momentum to give a quantity which should be momentum
independent, for different $p_T$ bins. As shown in
Figure~\ref{fig:dtheta}, $\delta\theta$ is the angular difference
between the reconstructed particle trajectory at the collision vertex
($x=0, y=0, z=z_{vtx}$) and at MuTR station~1. A
GEANT~\cite{GEANT:W5013} simulation of the PHENIX detector showed that
tracks which do not suffer a strong interaction in the pre-MuTR
absorber undergo angular deviations consistent with expectations based
on standard multiple scattering: $\sigma_{\delta\theta} \propto
\sqrt{x/X_0}/p$. The curves in each panel are fits to
$Cp\delta\theta\exp(-(p\delta\theta)^2/2(p\sigma_{\delta\theta})^2)$,
in which the normalization constant ($C$) is allowed to float, and
$p\sigma_{\delta\theta} = 130$\,rad$\cdot$MeV/c is given by GEANT and
is consistent with a simple estimate based on the radiation length of
the pre-MuTR absorber and the standard multiple scattering
formula~\cite{highland:1975,highland:1979,lynch:1991} ($x/X_0 \approx
48 \rightarrow p\theta_{space}^{rms} \approx (\sqrt{2})(13.6\,{\rm
MeV/c})(\sqrt{48})\,{\rm rad} = 133$\,rad$\cdot$MeV/c). The integral
beyond $3p\sigma_{\delta\theta}$ is $\approx 5\%$ and is largely due
to hadrons which have a strong interaction in the pre-MuTR absorber
and are still reconstructed as a muon candidate. Such tracks are
accounted for in the calculation of the punchthrough hadron yield, as
described below.

\subsubsection{Acceptance Restriction}
\label{sec:accrestrict}
We further restricted the acceptance of muon candidates for this
analysis in two ways:
\begin{enumerate}
\item We required tracks to pass through $\theta/\phi$ regions in
which the Monte Carlo detector response strictly agreed with the
measured response. This was established by agreement between the
number of data hits and Monte Carlo hits assigned to tracks in each
$\theta/\phi$ region of the detector.
\item We required tracks to lie within a pseudorapidity range,
  $1.5 < |\eta| < 1.8$, a region over which the acceptance depends
  only weakly on the collision $z_{vtx}$ location.
\end{enumerate}

\subsection{Acceptance and Efficiency}
\label{sec:acceff}

We factorized the acceptance and efficiency for tracks penetrating
to a particular MuID gap, $i$, into four components:
\begin{enumerate}
\item $\varepsilon^i_{\rm acc}$: the acceptance of a perfectly working
detector with the acceptance restrictions described above. This
quantity ($\approx 50\%$) is normalized to $2\pi\delta\eta$
($\delta\eta = 0.3$) and accounts for non-sensitive structural members
in between the cathode strip chambers and chamber regions removed
from consideration for the entirety of this analysis.
\item $\varepsilon^i_{\rm rec}$: the efficiency of reconstructing a
track that fell within the accepted region. This quantity is somewhat
low ($64\%$) due to detector problems in this first physics run that
have been subsequently resolved.
\item $\varepsilon^i_{\rm user}$: the efficiency for reconstructed
tracks to pass the cuts listed in Table~\ref{tab:roadcuts}.
\item $\varepsilon^i_{\rm trig}$: the efficiency of the MuID
trigger to fire in events with selected tracks.
\end{enumerate}

$\varepsilon^i_{\rm acc}$, $\varepsilon^i_{\rm rec}$, and
$\varepsilon^i_{\rm user}$ were evaluated with a GEANT simulation
using single muons thrown with a realistic $p_T$ spectrum into the
muon arms. The applied detector response incorporated measured
detector performance. Reductions in efficiency due to occupancy are
negligible in $p+p$ collisions. Run-to-run variations were ignored
since we selected runs in which the detector performance was similar
and stable. Efficiency values for tracks penetrating to MuID gap 4
were parameterized in terms of $z_{vtx}$ and $p_T$ and are listed in
Table~\ref{tab:efficiencies}. There are minor differences in these
parameterizations for particles with different charge sign.

\begin{table}[tbh]
\caption{Trigger, acceptance, track reconstruction and track selection
  efficiencies. Systematic errors for these quantities are given in
  Tables~\ref{tab:sigmaI} and~\ref{tab:sigmaP}.}
\label{tab:efficiencies}
\begin{ruledtabular}\begin{tabular}{cc}
Quantity & Value \\ \hline
$\varepsilon^{4,+}_{\rm acc}$                   & 
$0.51 \times (1 - 114\exp(-5.9 p_T)) \times (1 + 0.0015 z_{vtx})$\\ 
$\varepsilon^{4,-}_{\rm acc}$                   &
$0.50 \times (1 - 531\exp(-7.5 p_T)) \times (1 + 0.0013 z_{vtx})$\\ 
$\varepsilon^4_{\rm rec}$                       & 0.64 \\
$\varepsilon^{4,+}_{\rm user}$                   & 
$0.74 \times (1 - 0.0019 z_{vtx})$\\ 
$\varepsilon^{4,-}_{\rm user}$                   & 
$0.74 \times (1 - 0.0009 z_{vtx})$\\ 
$\varepsilon^3_{\rm scale}$                    & 0.66 \\
$\varepsilon^4_{\rm trig}$                  & 0.86 \\ 
$\varepsilon^3_{\rm trig}$                  & 0.97 \\ 
$\varepsilon_{BBC}^{c,\bar{c}\rightarrow\mu}$  & 0.75 \\ 
\end{tabular}\end{ruledtabular}
\end{table} 

We also determined the efficiencies for tracks which only penetrate to
MuID gap 3, since these are needed to obtain the yield of punchthrough
hadrons. These were found to scale from the efficiencies for tracks
penetrating to MuID gap 4: $\varepsilon^3_{\rm acc} \varepsilon^3_{\rm
rec} \varepsilon^3_{\rm user} = \varepsilon^3_{\rm scale} \times
\varepsilon^4_{\rm acc} \varepsilon^4_{\rm rec} \varepsilon^4_{\rm
user}$, where $\varepsilon^3_{\rm scale} = 0.66$.  $\varepsilon^3_{\rm
scale}$ is less than one because the MuID and the road reconstruction
algorithm are optimized for deeply penetrating particles. Particles
which do not penetrate to the last gap have poorer resolution matching
to MuTR tracks (due to reduced lever arm and smaller number of
hits) and are also more susceptible to MuID inefficiencies.

Trigger efficiencies, $\varepsilon^3_{\rm trig}$ and
$\varepsilon^4_{\rm trig}$, are also listed in
Table~\ref{tab:efficiencies}. These were evaluated using the {\em BBC}
data set, which did not require the MuID trigger to fire.
\begin{equation}
\label{eq:triggereff}
  \varepsilon^4_{\rm trig} = \frac
{(N_4 | {\it M1D}) \times S_{\it M1D}}
{(N_4 | {\it BBC}) \times S_{\it BBC}},
\end{equation}
where $N_4 | {\it M1D}$ is the number of selected tracks penetrating
to MuID gap 4 for events in which the {\em M1D} trigger fired, $S_{\it
M1D}$ is the scaledown factor applied to the {\em M1D} trigger, and
similarly for ${\it M1D} \rightarrow {\it BBC}$. $\varepsilon^3_{\rm
trig}$ was also evaluated according to
 Equation~\ref{eq:triggereff}, but with $N_4 \rightarrow N_3$, and
${\it M1D} \rightarrow {\it M1S}$.

Since both the {\em M1D} and {\em M1S} triggers required the {\em BBC}
trigger in coincidence with the MuID trigger we must also account for
the {\em BBC} trigger efficiency for events in which a reconstructed
muon is created via charm quark decay:
$\varepsilon_{BBC}^{c,\bar{c}\rightarrow\mu}$. The BBC efficiency was
evaluated for events in which a $J/\psi$ was produced in the muon arm
acceptance using PYTHIA+GEANT simulations~\cite{ppg017}. The BBC
efficiency was also evaluated for events in which $\pi^0$'s were
produced in the central arm acceptance~\cite{ppg024} using data
triggered without a BBC requirement. The BBC efficiency under both
conditions was found to have a similar value that we therefore adopt:
$\varepsilon_{BBC}^{c,\bar{c}\rightarrow\mu} = 0.75$.

Systematic errors for all acceptance and efficiency corrections are
discussed in Section~\ref{sec:syst} and listed in
Tables~\ref{tab:sigmaI} and~\ref{tab:sigmaP}.

\subsection{Inclusive Muon Candidates}
\label{sec:inclus}
\begin{figure*}[thb]
\includegraphics[width=0.48\linewidth]{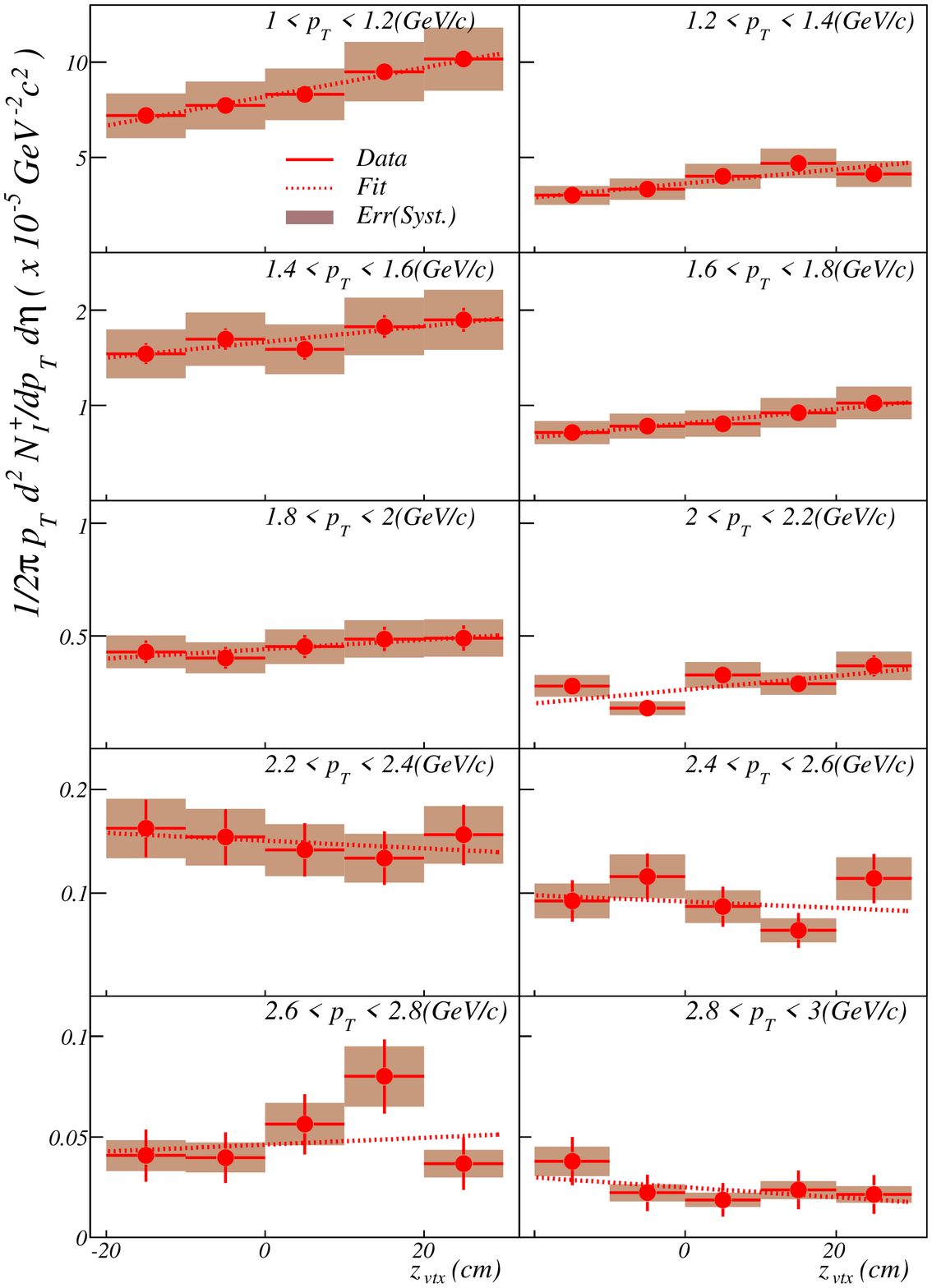}
\includegraphics[width=0.48\linewidth]{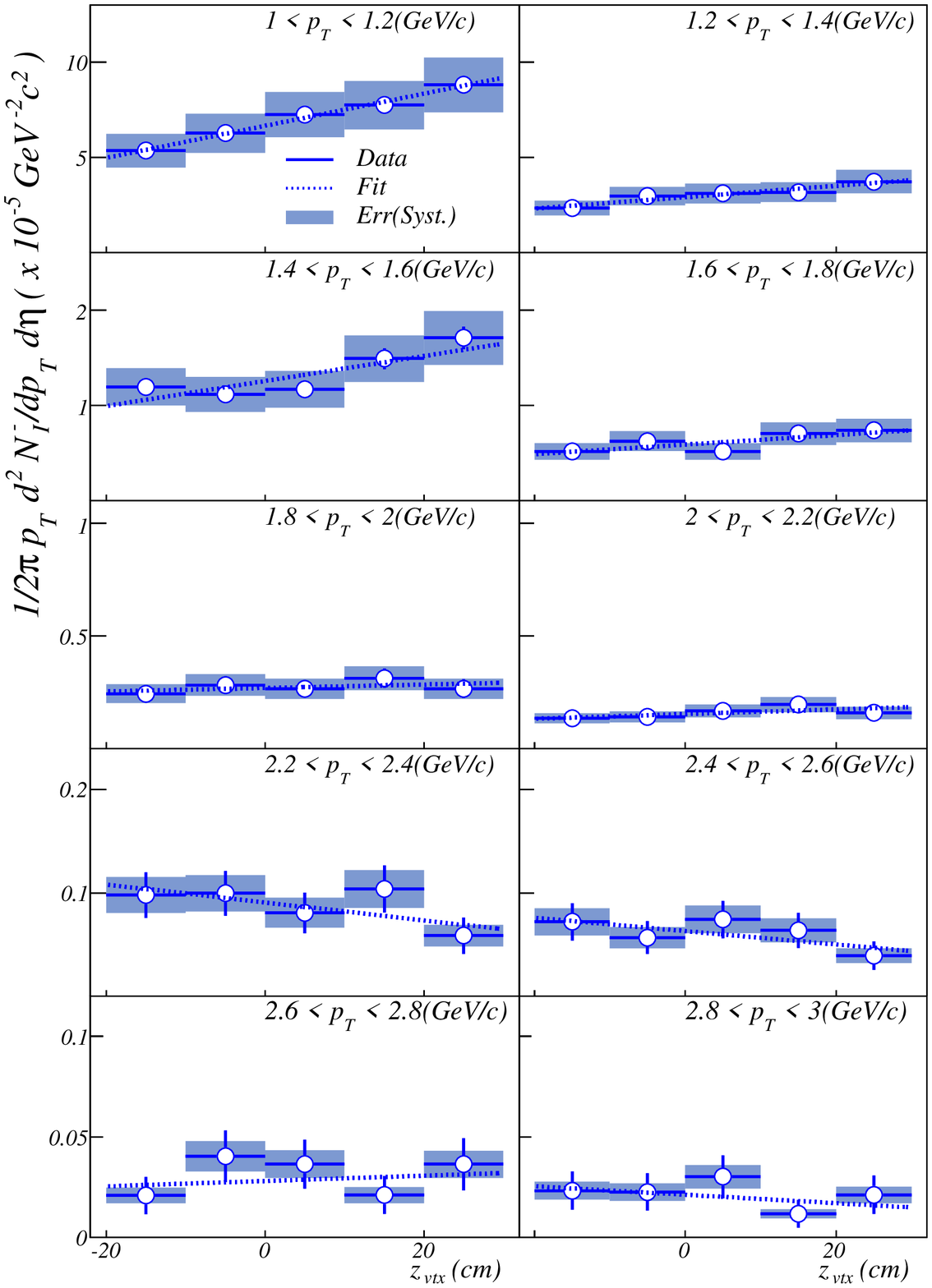}
\caption{Yield of (left) positively and (right) negatively charged
  inclusive muon candidates vs.~$z_{vtx}$ for different $p_T$
  bins. Fits shown use the functional form $a+bz_{vtx}$ to extract the
  contribution from hadron decay, as discussed in
  Section~\ref{sec:decay}. Error bars show statistical errors.  Shaded
  bands show systematic errors, as discussed in Section~\ref{sec:syst}
  and listed in Table~\ref{tab:sigmaI}.}
\label{fig:ni_plusminus}
\end{figure*}

We first form two sets of collision vertex ($z_{vtx}$) histograms with
10\,cm bins: one histogram for all interactions selected with the {\em
BBC} trigger, and a series of histograms for interactions selected with
the {\em M1D} trigger and having a good muon candidate within a $p_T$
bin ($1 < p_T < 3$\,GeV/c, 200\,MeV/c bins). The muon candidate
histograms are formed separately for each charge sign. Entries into
each histogram are scaled by the appropriate trigger scaledown
factor. The muon candidate histograms are divided by the minimum bias
histogram to give $N_I(p_T, z_{vtx})$, as shown in
Figure~\ref{fig:ni_plusminus}. Systematic errors shown in this figure
are discussed in Section~\ref{sec:syst} and listed in
Table~\ref{tab:sigmaI}.

\subsection{Hadron Generator}
\label{sec:cocktail}
In order to determine the contributions to the inclusive
muon yield from free-decay muons (Section~\ref{sec:decay}), and
punchthrough hadrons (Section~\ref{sec:punch}) we make
use of a data-driven hadron generator. The input for this generator is
obtained from PHENIX measurements in $\sqrt{s}=200$\,GeV $p+p$
collisions at $y=0$~\cite{ppg045,riabov:2005} using the following procedure:

\begin{enumerate}

\item $\pi^+$ and $\pi^-$ spectra at $y=0$ ($0<p_T<3.5$\,GeV/c) are
fit to a power law. We assume factorization in $y$ and $p_T$ and scale the spectra
fit at $y=0$ according to: 

$N^{\pi^\pm}_{y=1.65}(p_T) =
N^{\pi^\pm}_{y=0}(p_T)\exp(-(1.65^2/2\sigma_y^2))$,

with $\sigma_y = 2.5$. This factorization is observed both in
PYTHIA and in BRAHMS~\cite{Arsene:2004} data
measured at $y=1$ and $y=2.2$.

\item We use a similar procedure to obtain the charged kaon yield at
  $y=1.65$, but we need to first extrapolate the yield at $y=0$ beyond
  the current measurement limit ($p_T < 2$\,GeV/c).

We start by forming the isospin averaged $K/\pi$ ratio vs.~$p_T$ at
$y=0$. For $p_T<2$\,GeV/c we use charged particles,
$(K^++K^-)/(\pi^++\pi^-)$~\cite{ppg045}. We use neutral particles,
$K^0/\pi^0$, for $2 < p_T < 6.5$\,GeV/c~\cite{riabov:2005}. We then
fit this combined ratio to the form $f(p_T) =
A(1-B\exp(-Cp_T))$. Next, we normalize this function separately to the
$K^+/\pi^+$ and $K^-/\pi^-$ ratios for $p_T < 2$\,GeV/c. Finally, we
multiply by the corresponding charged pion spectrum to obtain
$N^{K^\pm}_{y=0}(p_T)$, our parameterization of the mid-rapidity
charged kaon $p_T$ spectra extending out to 3.5\,GeV/c.

As with pions, we need to extrapolate this parameterization of the
yield at $y=0$ to obtain the yield at $y=1.65$. One possibility is to
assume boost invariance of the $K/\pi$ ratio. However, PYTHIA gives a
slightly narrower rapidity distribution for kaons than for pions,
resulting in a kaon yield at $y=1.65$ that is only 85\% of that
predicted with the boost invariance assumption. We split the
difference between these two assumptions:

$N^{K^\pm}_{y=1.65}(p_T) = 92.5\%
N^{K^\pm}_{y=0}(p_T)\exp(-(1.65^2/2\sigma_y^2))$,

where, again, $\sigma_y = 2.5$. 

\item The $p$ and $\bar{p}$ spectra are assumed to have the same
shape as the pion spectra with normalization factors set to the
measured values at $y=0$, $p_T = 3$\,GeV/c (0.4 for $p/\pi^+$, 0.24 for
$\bar{p}/\pi^-$)~\cite{ppg045}. The exact form used for the $p,
\bar{p}$ spectra is unimportant. They obviously do not contribute to
the yield of decay muons and their contribution to the yield
of punchthrough hadrons is greatly suppressed due to their relatively
short nuclear interaction length.

\end{enumerate}

Systematic errors associated with this hadron generator are 
discussed in Section~\ref{sec:syst} and listed in Table~\ref{tab:sigmaD}.

\subsection{Free-Decay Muons}
\label{sec:decay}

\begin{figure*}[thb]
\includegraphics[width=0.8\linewidth]{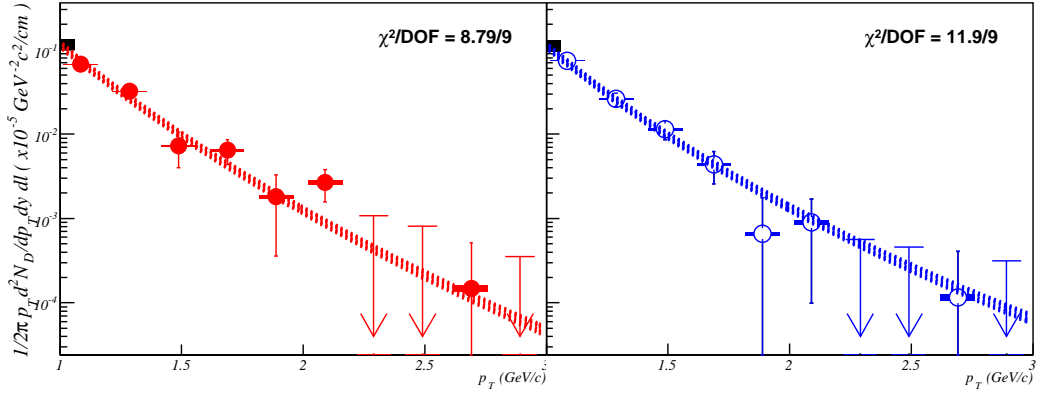}
\caption{Yield per unit length of (left) positively and (right)
  negatively charged free-decay muons. Points are the measured values
  determined by linear fits to the inclusive muon candidate yield
  (Figure~\ref{fig:ni_plusminus}). Error bars indicate statistical
  errors for those fits. $p_T$ bins with a negative (non-physical)
  slope in those fits are shown with a line at the 90\% C.L.U.L
  (statistical errors only) and an arrow pointing down. See
  Section~\ref{sec:decay} for details.  Dashed lines are the
  predictions for each sign from a data-driven hadron generator
  normalized to the measured points. The $\chi^2/dof$ for these fits
  are quoted. See Section~\ref{sec:cocktail} for details.  The width
  of the lines corresponds to $\sigma_{R_D}$ (see
  Table~\ref{tab:sigmaD}), the error on the ratio of free-decay muons
  to inclusive muon candidates.  Black bands at the left edge of each
  panel show the $p_T$-independent systematic error on the inclusive
  muon candidate yield. Shaded bands on each point show the systematic
  errors that affect the $p_T$ shape of the inclusive muon candidate
  spectrum. These last two systematic errors (Table~\ref{tab:sigmaI})
  need to be included in the total error budget for the yield of
  free-decay muons, $\sigma_{N_D}/N_D$ (Table~\ref{tab:sigmaD}), but
  are displayed separately since they are common to all components of
  the inclusive muon candidate yield, see Equations~\ref{eq:syserr0}
  and~\ref{eq:syserr}. Systematic errors are discussed in
  Section~\ref{sec:syst}.}
\label{fig:nd_plusminus}
\end{figure*}

In Figure~\ref{fig:ni_plusminus} one can clearly see the linear
dependence in the yield of inclusive muon candidates vs.~$z_{vtx}$ at
low transverse momentum ($p_T < 2$\,GeV/c). This dependence is due to
muons from the decay of abundant light hadrons ($\pi$'s and $K$'s)
prior to the first absorber material at $z_{abs} = -40$\,cm. We fit
these histograms with the function $a + bz_{vtx}$. After multiplying
by $dz/dl_{fp} = \cos(<\theta>) = 0.947$ the slope, $b$, and its fit
error give, respectively, the yield {\em per unit length of decay
flightpath} of muons from hadron decay, $dN_D(p_T)/dl_{fp}$, and the
statistical error on this quantity. Results are shown in
Figure~\ref{fig:nd_plusminus}. Systematic errors shown in this figure
are discussed in Section~\ref{sec:syst} and listed in
Table~\ref{tab:sigmaI}.

This procedure does not provide a quantitative measure of the decay
muon spectrum above $p_T \sim 2$\,GeV/c, even though a substantial
fraction of the inclusive muons are decay muons up to significantly
higher $p_T$. This is due to the fact that at high $p_T$ the decay
slopes decrease (Lorentz time dilation) as do the statistics, both of
which make it more difficult to quantify the decay component
directly. In order to extend our estimate of decay muons to higher
$p_T$ we use our hadron generator, described in
Section~\ref{sec:cocktail}. We simulate the decay of hadrons into the
muon arms and obtain predicted $p_T$ spectra (per unit length) of
muons from hadron decay separately for each charge sign. We then
normalize these predicted spectra to the measured spectra. The
normalized predicted spectra are shown as the dashed lines in
Figure~\ref{fig:nd_plusminus}. The predicted spectral shape agrees
with the data where we have a statistically significant
measurement. The absolute normalization of the prediction is within
7\% of the measured value, easily consistent within errors.

We obtain $N_D(p_T, z_{vtx})$ from the product of $dN_D(p_T)/dl_{fp}$
and the average value of the decay flightpath, $l_{fp} = \lambda_D +
|z_{vtx} - z_{abs}|/\cos(\theta)$, for each $z_{vtx}$ bin.

\subsection{Punchthrough Hadrons}
\label{sec:punch}
A hadron penetrating to MuID gap 4 is impossible to distinguish
from a muon. However, we can cleanly identify hadrons in shallower
gaps and then extrapolate their yield to obtain the yield of
punchthrough hadrons in MuID gap 4.

\begin{figure}[t]
\includegraphics[width=1.05\linewidth]{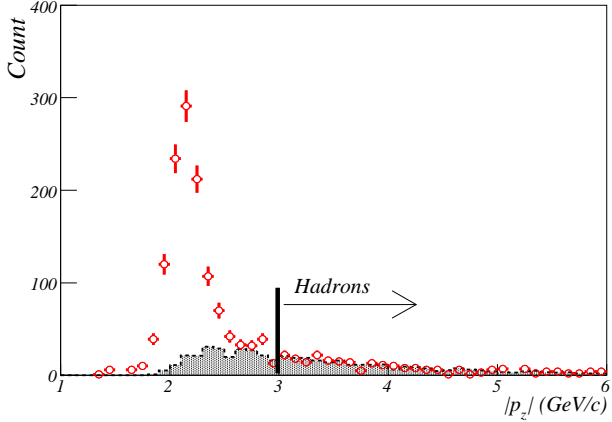}
\caption{Points (measured data) show the longitudinal momentum,
measured at the vertex ($p_z^{vtx}$), of particles that stop in MuID
gap 3. The sharp peak is due to muons which range out in the absorber
plate between gaps 3 and 4. The histogram (Monte Carlo) shows the
longitudinal momentum of all particles that stop in MuID gap 3 and
do not decay before the pre-MuTR absorber. The Monte Carlo is
normalized to the data for $p_z^{vtx}>3$\,GeV/c. Particles beyond the
peak form a relatively pure sample of hadrons.}
\label{fig:pzgap3}
\end{figure}

\begin{figure*}[tbh]
\includegraphics[width=0.8\linewidth]{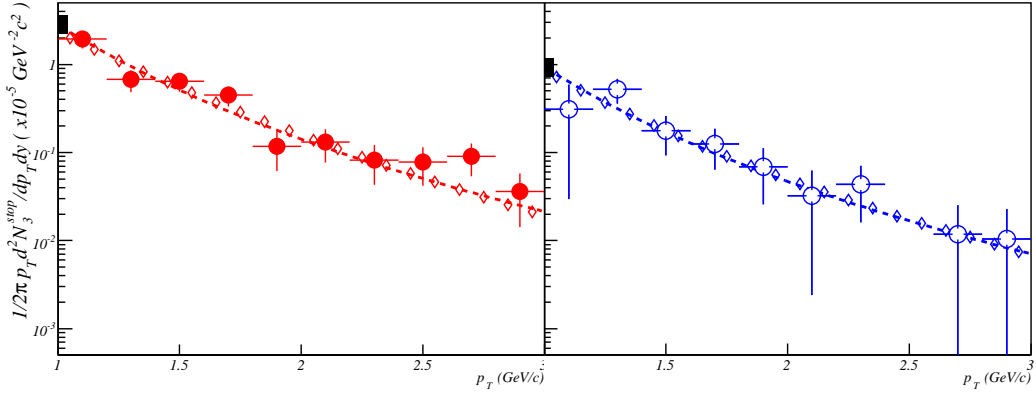}
\caption{Points show the $p_T$ spectrum of (left) positively and
  (right) negatively charged hadrons stopping in MuID gap 3 (``gap 3
  exclusive yield'') with statistical errors. Open diamonds show a
  power-law fit to the data, effectively a smoothed version of the
  data. Dashed lines are fits to the smoothed data using the hadron
  generator (normalized by the measured free-decay spectrum, as shown
  in Figure~\ref{fig:nd_plusminus}) and Equations~\ref{eq:h3splus}
  and~\ref{eq:h3sminus} to obtain values for the species-dependent
  nuclear interaction lengths, $\lambda^i(p_T)$.}
\label{fig:pt_punchthrough}
\end{figure*}

Figure~\ref{fig:pzgap3} shows the longitudinal momentum ($p_z$, the
momentum projected onto the beam axis) distribution of particles that
{\em stop} in MuID gap 3. The sharp peak at $p_z \approx 2.2$\,GeV/c
corresponds to charged particles which stopped because they ranged out
in the absorber plate between gaps 3 and 4 (this includes both muons
and also 
hadrons which only suffered ionization energy loss.) The
width of this peak is due to the 20\,cm (11.4\,$X_0$) absorber
thickness between MuID gaps 3 and 4, and energy-loss fluctuations in
all the preceding absorber layers. Particles at momenta beyond the
peak ($p_z > 3$\,GeV/c) form a relatively pure sample of hadrons, with
only a small contamination due to inefficiencies in MuID gap 4 and
particles with mis-reconstructed momentum values. After correcting for
acceptance and efficiency we use these particles to obtain the $p_T$
spectrum for the ``gap 3 exclusive yield,'' as shown in
Figure~\ref{fig:pt_punchthrough}. Note: we use data from the {\em M1S}
trigger sample since the {\em M1D} sample required a hit in MuID gap
4, which would bias this measurement.


In order to extrapolate this measured spectrum for hadrons stopping in
MuID gap 3 to the spectrum of punchthrough hadrons which penetrate to
MuID gap 4 we start by assuming exponential absorption of hadrons
entering the muon arm absorber material.  With this assumption we
obtain an expression for the ``gap 3 inclusive yield'', those hadrons
that reach {\em at least} MuID gap 3:
\begin{equation}
\label{eq:h3}
N^i_3(p_T,\theta) = N_{vtx}^i(p_T,\theta) \exp(-L_3(\theta)/\lambda^i(p_T)),
\end{equation}
where $i$ indicates the contributing hadron species ($\pi^{\pm},
K^{\pm}, p, \bar{p}$), $N^i_{vtx}(p_T,\theta)$ is the yield at the
vertex of the $i^{\rm th}$ species, $L_3(\theta)$ is the amount of
absorber material traversed to reach MuID gap 3, and
$\lambda^i(p_T,\theta)$ is the $p_T$-dependent nuclear interaction
length of the $i^{\rm th}$ species.  We can write a similar expression
for the punchthrough hadron yield:
\begin{equation}
\label{eq:h4}
N_P^i(p_T,\theta) = N_{vtx}^i(p_T,\theta) \exp(-L_4(\theta)/\lambda^i(p_T)),
\end{equation}
where $L_4(\theta)$ is the amount of absorber material traversed to reach
MuID gap 4.

By taking the difference between these two equations we obtain an
expression for the gap 3 exclusive yield:
\begin{widetext}
\begin{eqnarray}
\label{eq:h3s}
N_3^{i,stop}(p_T,\theta) & = & N_3^i(p_T,\theta) - N_P^i(p_T,\theta) \nonumber \\
      & = & N_{vtx}^i(p_T,\theta)
      \exp(-L_3(\theta)/\lambda^i(p_T)) \nonumber \\
\times  (1 - \exp((L_3(\theta) - L_4(\theta))/\lambda^i(p_T)))
\end{eqnarray}

In our measurement we cannot identify the species comprising the gap 3
exclusive yield, but we do know their charge sign. As a result,
Equation~\ref{eq:h3s} can be rewritten as two equations with six
unknowns for each $p_T$ bin:
\begin{eqnarray}
\label{eq:h3splus}
N_3^{+,stop}(p_T,\theta) & = & N_3^+(p_T,\theta) - N_P^+(p_T,\theta) \nonumber \\
      & = & \sum_{i = \pi^+, K^+, p} N_{vtx}^i(p_T,\theta)
      \exp(-L_3(\theta)/\lambda^i(p_T))  \nonumber \\
\times (1 - \exp((L_3(\theta) - L_4(\theta))/\lambda^i(p_T))),
\end{eqnarray}
\begin{eqnarray}
\label{eq:h3sminus}
N_3^{-,stop}(p_T,\theta) & = & N_3^-(p_T,\theta) - N_P^-(p_T,\theta) \nonumber \\
      & = & \sum_{i = \pi^-, K^-, \bar{p}} N_{vtx}^i(p_T,\theta)
      \exp(-L_3(\theta)/\lambda^i(p_T)) \nonumber \\ 
\times (1 - \exp((L_3(\theta) - L_4(\theta))/\lambda^i(p_T))).
\end{eqnarray}
\end{widetext}

Based on measured cross sections for various species~\cite{pdg}, we
chose to reduce the number of unknowns with the following assumption:
\begin{eqnarray}
\lambda_{K^+} & = & \lambda_{long}, \nonumber \\
\lambda_p = \lambda_{\pi^+} = \lambda_{\pi^-} = \lambda_{K^-}  & = &
\lambda_{short}, \nonumber \\
\lambda_{\bar{p}} & = & 0. \nonumber
\end{eqnarray}
We further assume that $\lambda_{short}$ and $\lambda_{long}$ have the
form $a+b(p_T{\rm [GeV/c]} - 1)$.

We effectively smoothed the gap 3 exclusive yield for each sign by
fitting the measured values to a power law. Using
$N_{vtx}^i(p_T,\theta)$ from the hadron generator (normalized to the
free-decay muon spectrum, as described in Section~\ref{sec:decay}) and
known values for $L_{3,4}(\theta)$, we fit Equations~\ref{eq:h3splus}
and~\ref{eq:h3sminus} to the smoothed gap 3 exclusive yield for each
sign to obtain:
\begin{eqnarray}
\lambda_{short} & = & 19.0 + 2.2(p_T {\rm [GeV/c]} - 1)\,{\rm cm,~and}
\nonumber \\
\lambda_{long}  & = & 25.9 + 4.4(p_T {\rm [GeV/c]} - 1)\,{\rm cm.} \nonumber 
\end{eqnarray}
Results of these fits are shown in Figure~\ref{fig:pt_punchthrough}.

With these values for $\lambda_i(p_T)$ and the hadron generator
input spectra, we could directly apply Equation~\ref{eq:h4} to obtain the
final punchtrough spectra. However, we made one further correction,
described below, after finding that our assumption of exponential
absorption does not hold when applied to GEANT simulations of the
punchthrough process.

Using our GEANT-based PHENIX simulation program, we generated data
sets with both the FLUKA~\cite{FLUKA} and GHEISHA~\cite{GHEISHA}
hadronic interaction packages. Input spectra for these data sets were
given by our decay hadron generator. We selected all particles which
did not decay before the pre-MuTR absorber. ``Truth'' values for the
punchthrough and gap 3 exclusive yields were obtained by splitting
those particles based on the absence (gap 3 exclusive) or presence
(punchthrough) of associated charged particles with $E_4 > 100$\,MeV
in MuID gap 4. We varied $E_4$ from $50-300$\,MeV and saw no
significant change in the results.

Using the known input spectra, known values for $L_{3,4}(\theta)$, and
truth values for gap 3 exclusive yield, we applied
Equation~\ref{eq:h3s} to the Monte Carlo data sets to obtain
$\lambda_i(p_T)$. Due to statistical limitations we integrated our
results over $\theta$ and into two $p_T$ bins: $1 < p_T < 2$\,GeV/c
and $p_T > 2$\,GeV/c. Values extracted for $\lambda^i(p_T)$ for the
different hadronic interaction packages are
listed in Table~\ref{tab:lambda}. These values are consistent with
those found for our measured data, listed above.

\begin{table}[tbh]
\caption{Nuclear interaction lengths, $\lambda^i(p_T)$, for different
  particle species and $p_T$ bins (in GeV/c) for FLUKA and
  GHEISHA. Statistical errors on these values are $\approx 2$\,mm.}
\label{tab:lambda}
\begin{ruledtabular}\begin{tabular}{ccccc}
& \multicolumn{2}{c}{FLUKA} & \multicolumn{2}{c}{GHEISHA}\\
& \multicolumn{2}{c}{$\lambda^i(p_T)$ [cm]} &
  \multicolumn{2}{c}{$\lambda^i(p_T)$ [cm]}               \\ \cline{2-5}
Species & $1 < p_T < 2$ & $p_T > 2$ 
        & $1 < p_T < 2$ & $p_T > 2$ \\ \hline
$\pi^+$ & 19.6 & 24.5 & 16.0 & 21.1 \\
$\pi^-$ & 19.4 & 24.8 & 15.0 & 19.3 \\
$K^+$   & 24.4 & 29.6 & 24.9 & 30.8 \\
$K^-$   & 20.5 & 24.2 & 17.2 & 21.2 \\ 
\end{tabular}\end{ruledtabular}
\end{table} 

\begin{table}[tbh]
\caption{Ratios, $R_{N_P^i(p_T)}$, of truth values for the
  punchthrough hadron yield to those predicted assuming exponential
  hadron absorption for different particle species and $p_T$ bins (in
  GeV/c), for FLUKA and GHEISHA. Average values of the ratios for the
  two different hadronic interaction packages, $\langle R
  \rangle^i(p_T)$, are smoothed across the $p_T$ bin at 2\,GeV/c to
  obtain correction factors for the exponential absorption
  model. Statistical errors on these quantities are
  $\approx$\,10\%. The maximum fractional difference in the ratios for
  the two different packages (32\%) is incorporated into the
  systematic error estimate, as shown in Table~\ref{tab:sigmaP}.}
\label{tab:h4ratio}
\begin{ruledtabular}\begin{tabular}{cccc}

Species & $1 < p_T < 2$ & $p_T > 2$ & Description \\ \hline

$\pi^+$ & 0.76 & 0.86 & \\
$\pi^-$ & 0.91 & 0.75 & \\
$K^+$   & 0.91 & 1.00 & $R^{FLUKA}_{N_p^i(p_T)}$ \\
$K^-$   & 1.17 & 1.06 & \\ \hline

$\pi^+$ & 1.48 & 1.04 & \\
$\pi^-$ & 1.47 & 1.09 & \\
$K^+$   & 1.31 & 1.07 & $R^{GHEISHA}_{N_p^i(p_T)}$\\
$K^-$   & 2.21 & 1.69 & \\ \hline

$\pi^+$ & 1.12 & 0.95 & \\
$\pi^-$ & 1.19 & 0.92 & \\
$K^+$   & 1.11 & 1.04 & $\langle R \rangle^i(p_T)$\\
$K^-$   & 1.67 & 1.38 & \\ \hline

$\pi^+$ & 32\% & 10\% & \\
$\pi^-$ & 24\% & 18\% & \\
$K^+$   & 18\% & 3\%  & $\delta R_{N_p^i(p_T)}/C^i(p_T)$ \\
$K^-$   & 32\% & 22\% & \\

\end{tabular}\end{ruledtabular}
\end{table}

\begin{figure}[tbh]
\includegraphics[width=0.85\linewidth]{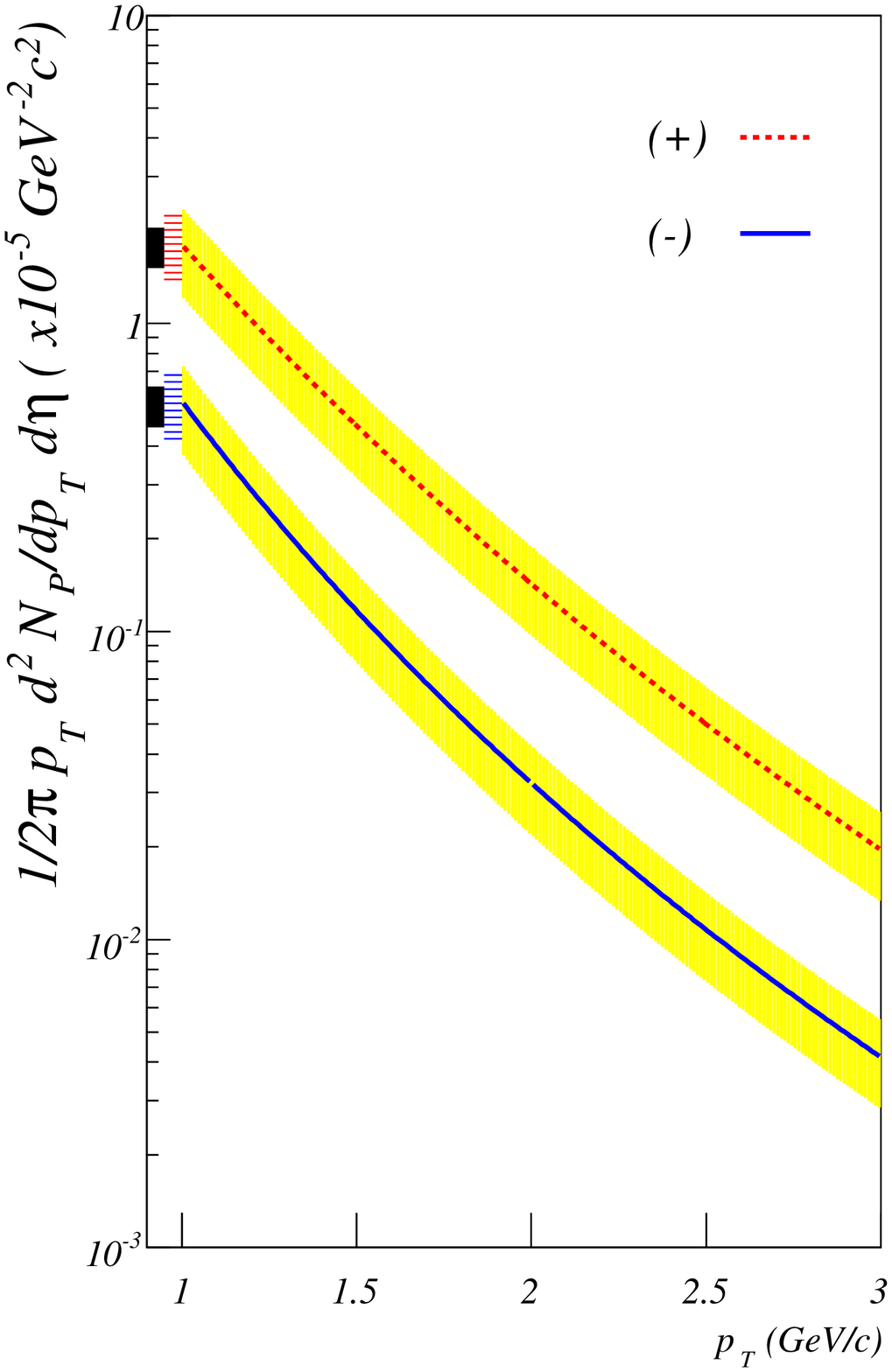}
\caption{Yield of positively (dotted line) and negatively (solid line)
  charged hadrons which penetrate to MuID gap 4. The curves are
  obtained from Equation~\ref{eq:h4corr}, as described above. Solid
  bands on the $y$-axis show the relative normalization error on the
  inclusive muon candidate yield, $\sigma^{\rm norm}_{N_I}/N_I$, see
  Table~\ref{tab:sigmaI}. This needs to be included in the total error
  budget for the yield of punchthrough hadrons, $\sigma_{N_P}/N_P$
  (Table~\ref{tab:sigmaP}), but is displayed separately since it is
  common to all components of the inclusive muon candidate yield, see
  Equations~\ref{eq:syserr0} and~\ref{eq:syserr}.  Hatched bands on
  the $y$-axis show $\sigma^{\rm norm}_{R_P}$ (see
  Table~\ref{tab:sigmaP}), the normalization error on the ratio of
  punchthrough hadrons to inclusive muon candidates. The relative
  fraction of positive and negative punchthrough hadrons can move up
  and down together by this amount.  Shaded bands around the extracted
  punchthrough hadron yield show the systematic errors on
  $\sigma^{p_T}_{R_P}$ which can affect the $p_T$ shape of the
  relative fraction of positive and negative punchthrough hadrons
  (positives and negatives can move independently). These are
  dominated by differences in the applicability of the simple
  exponential absorption model observed for FLUKA and
  GHEISHA. Systematic errors are listed in Tables~\ref{tab:sigmaI}
  and~\ref{tab:sigmaP} and discussed in Section~\ref{sec:syst}.}
\label{fig:np_plusminus}
\end{figure}

Inserting these values for $\lambda^i(p_T)$ into Equation~\ref{eq:h4}
we obtained a prediction for $N_P^i(p_T, \theta)$. Ratios of the truth
values and predicted values for the punchthrough yield
($R_{N_P^i(p_T)}$) for the different hadronic interaction packages are
listed in Table~\ref{tab:h4ratio}. One can see that these ratios
deviate significantly from unity and that the two hadronic interaction
packages disagree on the direction of the deviation: the exponential
absorption model tends to overpredict the punchthrough yield for FLUKA
($R^{FLUKA}_{N_P^i(p_T)} < 1$) and underpredict it for GHEISHA
($R^{GHEISHA}_{N_P^i(p_T)} > 1$).

Relatively little data exists in the relevant momentum range that
would allow us to conclude which, if either, of the hadronic
interaction package is correct. Measurements by RD10 and
RD45~\cite{RD45} of the penetration depth of identified hadrons found
that GHEISHA did well for protons and FLUKA did not. But neither did
well for pions and no data exists for kaons. Furthermore, the results
are sensitive to the definition of a penetrating particle: For RD10/45
an incoming particle with any associated charged particles in the
$120\times120$\,cm$^2$ detector area at a particular depth was defined
to have penetrated to that depth. In our measurement we reconstruct
particle trajectories and MuID hits are not associated with a road
unless they are within a narrow search window surrounding the projected
trajectory. Thus we are relatively insensitive to the leakage of a
showering hadron.

As a result of these uncertainties on the applicability of our
exponential absorption model we incorporate a species and $p_T$-dependent
correction factor to Equation~\ref{eq:h4}:
\begin{widetext}
\begin{equation}
\label{eq:h4corr}
N_P^i(p_T, \theta) = C^i(p_T)N_{vtx}^i(p_T,\theta) \exp(-L_4(\theta)/\lambda^i(p_T)).
\end{equation}

The correction factors for pions and kaons are obtained from the
average of the values of $R_{N_P^i(p_T)}$ for the two packages,
$\langle R \rangle^i(p_T) = (R^{FLUKA}_{N_P^i(p_T)} +
R^{GHEISHA}_{N_P^i(p_T)})/2$, which are listed in
Table~\ref{tab:h4ratio}. The values of $\langle R \rangle^i(p_T)$ for
a given species are not the same for the different $p_T$ bins, so we
assume the values are valid at the average $p_T$ of each bin ($p_T =
1.25$\,GeV/c and 2.31\,GeV/c respectively) and use a linear
extrapolation in $p_T$ to obtain the final correction factors:
\begin{equation}
C^i(p_T) = \langle R \rangle^i(1<p_T<2{\rm GeV/c}) + 
           (\langle R \rangle^i(p_T>2{\rm GeV/c}) - \langle R \rangle^i(1<p_T<2{\rm GeV/c}))  
           \frac{p_T[{\rm GeV/c}] - 1.25}{2.31 - 1.25}
\end{equation}
\end{widetext}
We assume that $p$'s and $\bar{p}$'s have the same correction factors
as the corresponding sign pions. Since $p$'s and $\bar{p}$'s make only
small contributions to the punchthrough hadrons this simplifying
assumption has little consequence. We incorporate the maximum
fractional difference in the ratios for the two packages (32\%) into
our systematic error estimate, as listed in Table~\ref{tab:sigmaP}.

We use Equation~\ref{eq:h4corr}, with the tabulated correction
factors, particle yields at the vertex given by our normalized hadron
generator, the known value of $L_4(\theta)$, and the values for
$\lambda^i(p_T)$ determined from the measured gap 3 exclusive yield,
to obtain the $p_T$ spectrum of punchthrough hadrons, $N_P(p_T)$, as
shown in Figure~\ref{fig:np_plusminus}. Systematic errors shown in
this figure are discussed in Section~\ref{sec:syst} and listed in
Table~\ref{tab:sigmaP}. We multiply $N_P(p_T)$ by the fraction of the
accepted $z_{vtx}$ range represented by each $z_{vtx}$ bin to finally
obtain $N_P(p_T, z_{vtx})$.

\subsection{Background Tracks}
\label{sec:back}
The main source of tracks which are not accounted for in the yield of
punchthrough hadrons and free-decay muons, and which are not due to
vertex-independent muons, are light hadrons which penetrate through
the pre-MuTR absorber, decay into a muon, and are still reconstructed
as a valid track.

A simulation of single pions thrown into the muon arm acceptance shows
that the number of hadrons which decay after the pre-MuTR absorber and
penetrate to MuID gap 4 is only 5-10\% (increasing with increasing
$p_T$) of the $z_{vtx}$-averaged number of free-decay muons, $N_D(p_T,
z_{vtx}=0)$. This ratio will be suppressed by the fact that tracks which
decay are less likely to be reconstructed successfully. It is further
suppressed by our punchthrough calculation procedure: the number of
such tracks which stop in MuID gap 3 is roughly half the number that
penetrate to gap 4; these will be counted in our calculation of the
punchthrough hadron yield. 

We express our estimate for the yield of background tracks not
otherwise accounted for as $N_B(p_T) = 5\% \times N_D(p_T,
z_{vtx}=0)$. The systematic uncertainty assigned to this quantity,
$\pm 5\% \times N_D(p_T, z_{vtx}=0)$, covers the extreme possibilities
that the $N_B/N_D$ is unsuppressed or fully suppressed by
reconstruction and punchthrough procedures, as described above.

This estimate was verified in a simulation of $\pi^-$'s and
$K^-$'s which were thrown into the muon arm acceptance and fully
reconstructed. The reconstructed track information, together with the
Monte Carlo truth information, allows us to eliminate uncertainties
due to mis-reconstruction of the track $p_T$ and due to determination
of whether a track which penetrated to the last gap did so in a
reconstructible fashion. Systematic errors on this estimate are
discussed in Section~\ref{sec:syst}.


\subsection{Vertex-Independent Muons}
\label{sec:heavy}

Figure~\ref{fig:ni_plusminusdecomposed} shows the yield of inclusive
muon candidates, $N_I(p_T, z_{vtx})$, with contributions from
individual components (free-decay muons, punchthrough hadrons, and
background tracks) shown as well as their sum. The vertex-independent
muons can be seen as the clear excess above the calculated background
sources. The systematic error bands shown on the component sums are
discussed in Section~\ref{sec:syst} and listed in
Tables~\ref{tab:sigmaD} and~\ref{tab:sigmaP}.

We obtain the yield of vertex-independent muons by applying
Equation~\ref{eq:muonsources} in each $p_T$ bin, subtracting the
hadronic contributions from the inclusive muon candidate yield, and
averaging over $z_{vtx}$ bins. This yield is shown, before averaging
over $z_{vtx}$ to demonstrate the expected vertex independence, in
Figure~\ref{fig:ni_plusminussubtracted}.

We make one final correction for momentum scale. The observed mass of
the $J/\psi$, reconstructed with the same code and in the same data
set, is higher than the nominal value by $\approx 100\,{\rm MeV}
(3\%)$\cite{ppg017}. However, in a higher statistics data set the
momentum scale accuracy is verified to within 1\% by our observation
of the accepted value for the mass of the
$J/\psi$~\cite{ppg038}. Also, the peak observed in the longitudinal
momentum distribution of particles stopping in MuID gap 3 (see
Fig.~\ref{fig:pzgap3}) is within $0.5\%$ of the predicted value. We
therefore assume that the momentum scale is high by $1.5\%$ (splitting
the difference between 0 and 3\%). This results in a momentum scale
correction factor to the prompt muon yield of $0.94 + 0.987 \times
(p_T{\rm [GeV/c]}-1)$. We assume a 100\% systematic error on this
correction factor, as shown in Tab.~\ref{tab:sigmaI}.


Finite momentum resolution can cause a similar effect. Contributions
from energy loss fluctuations, multiple scattering and chamber
resolution combine to give $\delta p/p \approx 5\%$ for the momentum
range used in this analysis. Finite resolution, combined with an
exponentially falling spectrum, artificially hardens the measured
spectrum. For $1 < p_T < 3$\,GeV/c this hardening increases the
normalization of the yield by $\approx 3.7$\%. However, this is
accounted for in our efficiency determination since
we use a realistic $p_T$ spectrum as input. Therefore we apply no
explicit correction for this effect.

The final values for the vertex-independent muon cross section,
obtained from Equation~\ref{eq:ptspectrum}, are shown in
Figure~\ref{fig:nprompt_plusminus}. Points in this figure have been
placed at the average $p_T$ value of the bin contents to account for
bin shifting in the steeply falling distributions. Systematic errors
shown in this figure are discussed in Section~\ref{sec:syst} and
listed in Tables~\ref{tab:sigmaI}~-~\ref{tab:sigmaP}.

\subsection{Systematic Errors}
\label{sec:syst}

\begin{figure*}[thb]
\includegraphics[width=0.4\linewidth]{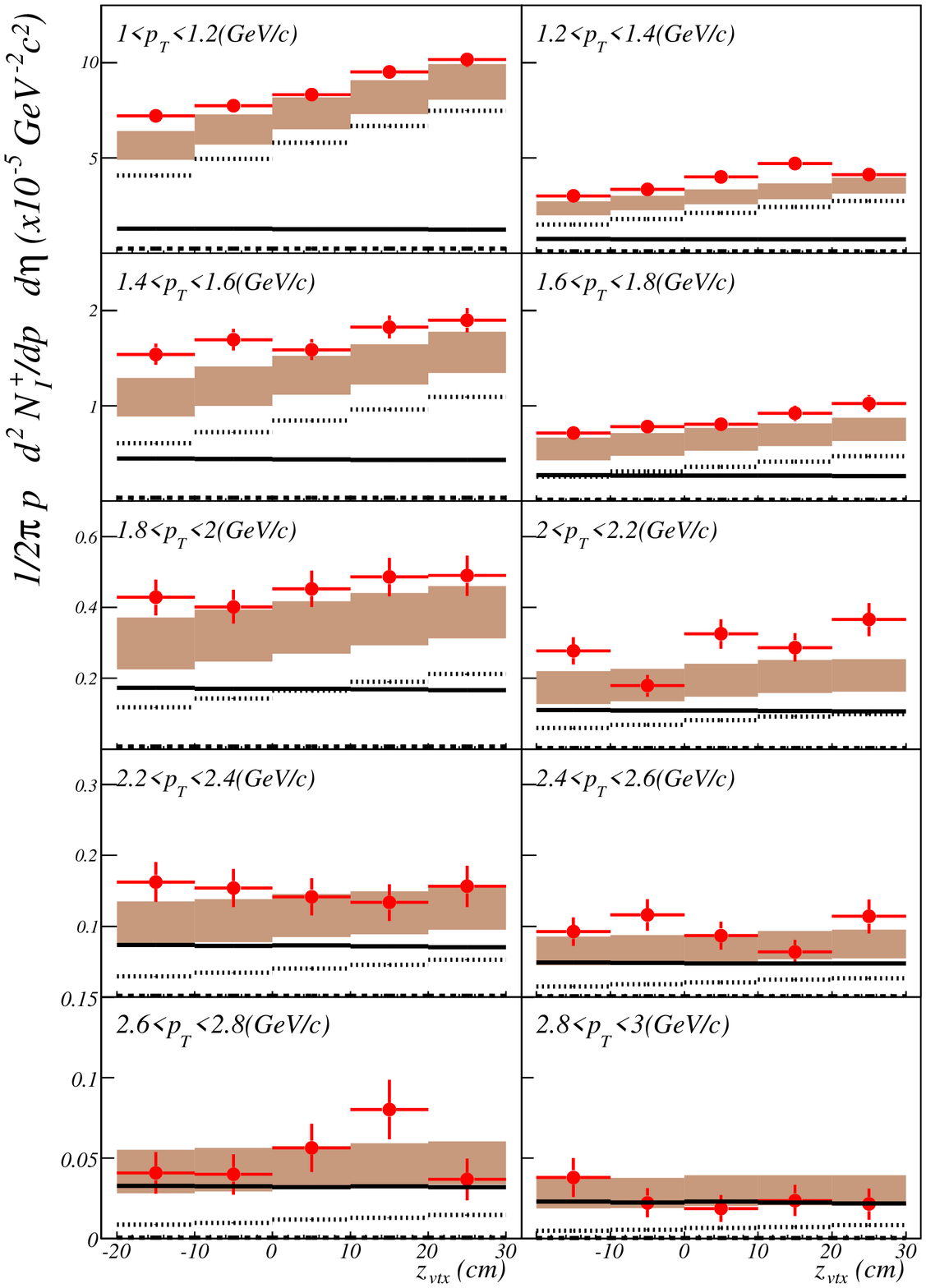}
\includegraphics[width=0.4\linewidth]{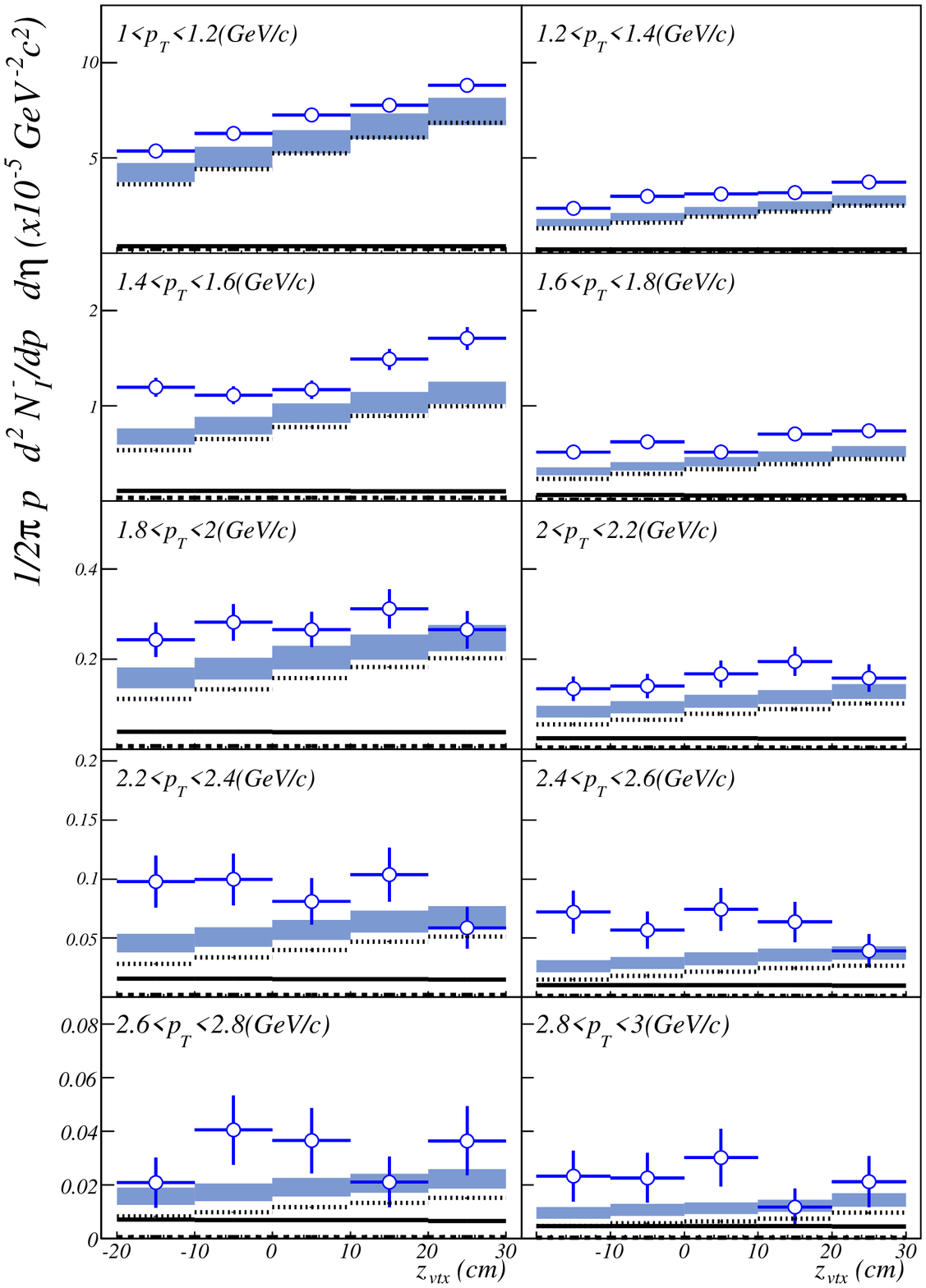}
\caption{Points show the yield of (left) positively and (right)
  negatively charged inclusive muon candidates vs.~$z_{vtx}$ for
  different $p_T$ bins with statistical errors. Dotted, solid and
  dashed lines show contributions from decay muons, punchthrough
  hadrons and background tracks, respectively. Shaded bands show the
  systematic error around the sum of these components, as listed in
  Tables~\ref{tab:sigmaI}-\ref{tab:sigmaP} and discussed in
  Section~\ref{sec:syst}.}
\label{fig:ni_plusminusdecomposed}
\end{figure*}

\begin{figure*}[thb]
\includegraphics[width=0.4\linewidth]{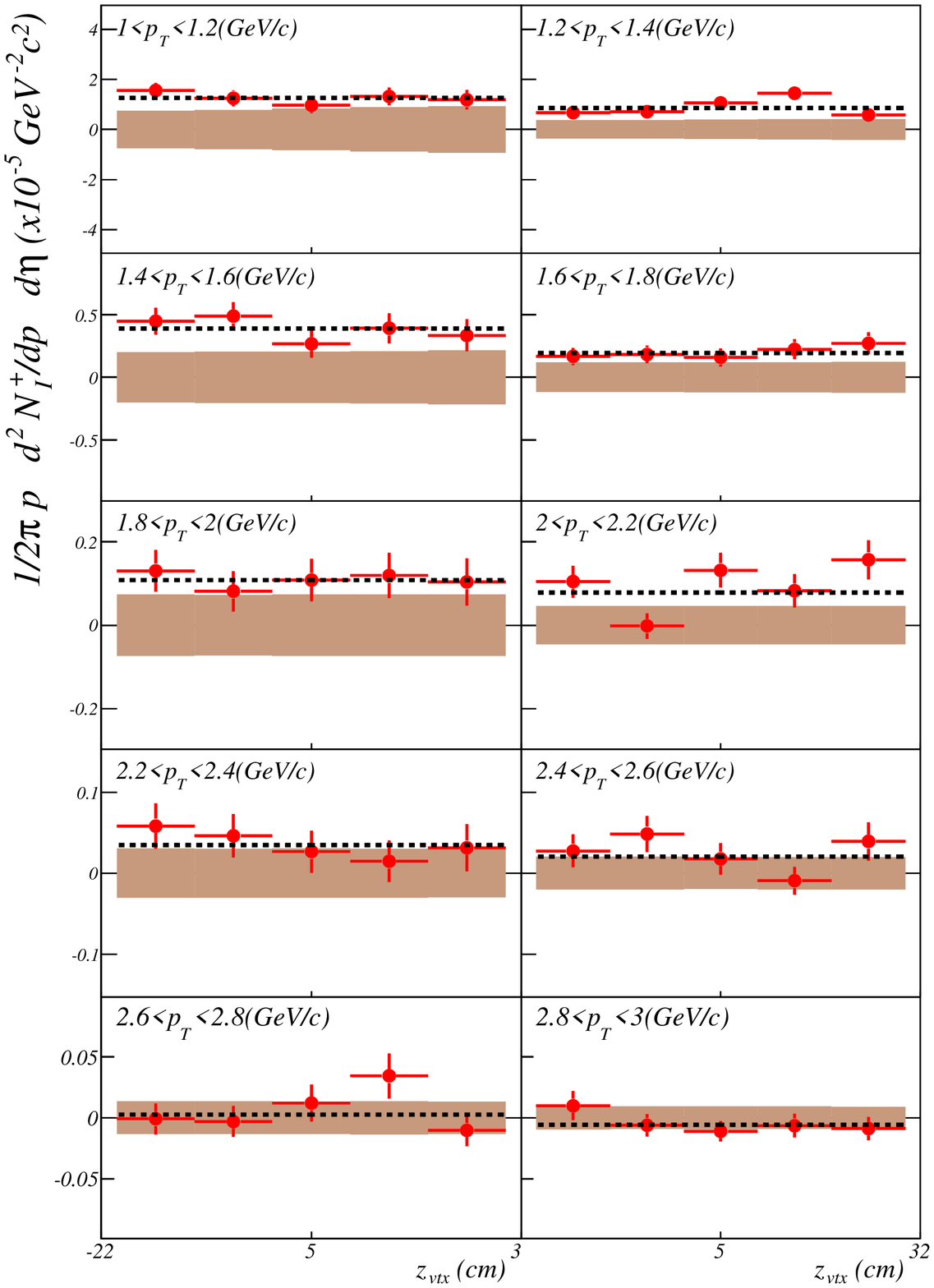}
\includegraphics[width=0.4\linewidth]{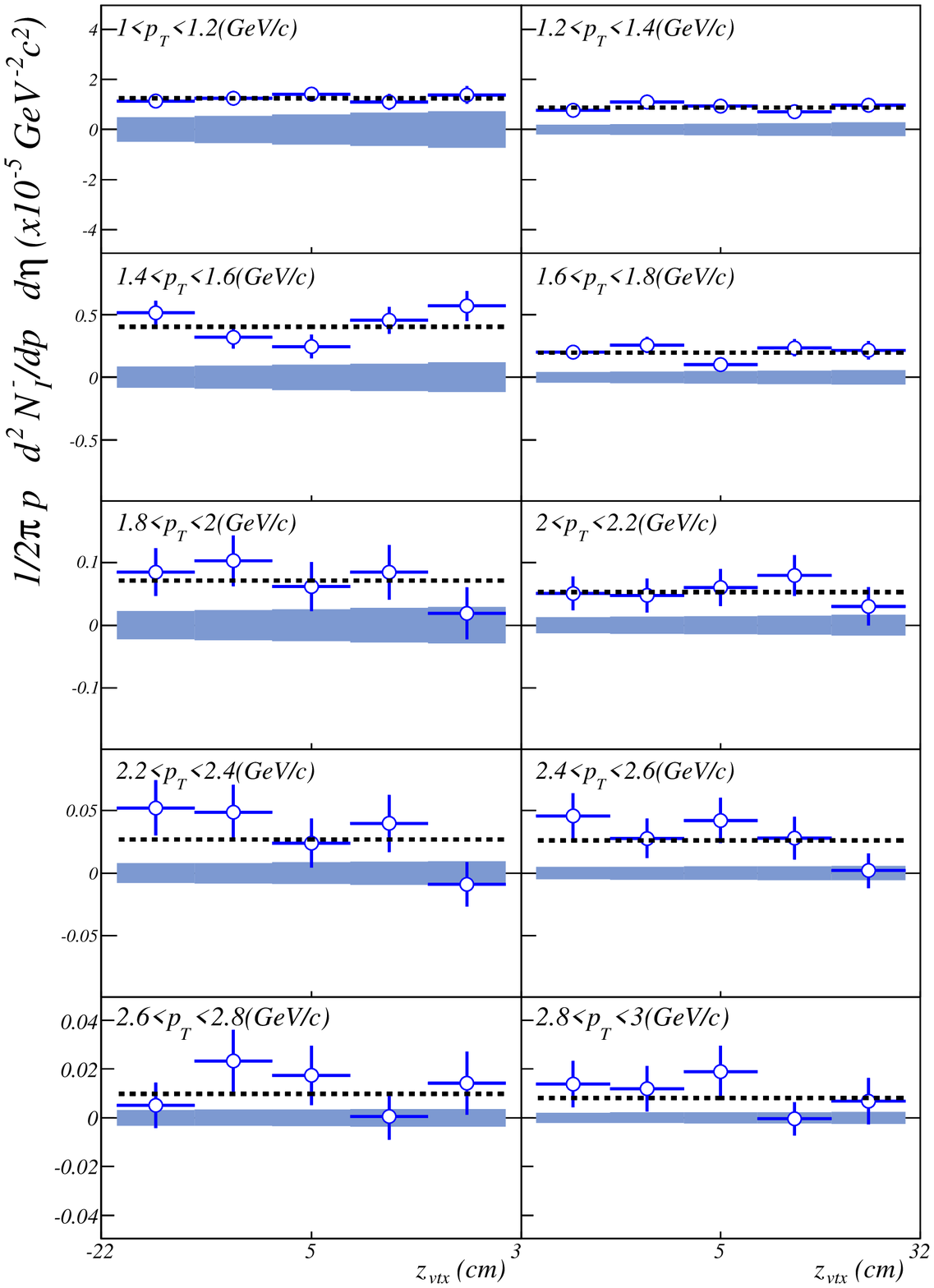}
\caption{Points show the yield of (left) positively and (right)
  negatively charged vertex-independent muons vs.~$z_{vtx}$ for
  different $p_T$ bins with statistical errors. The dashed lines show
  the yield for each $p_T$ bin, averaged over $z_{vtx}$. The shaded
  bands around 0 show the systematic error on the sum of the
  contributions to the inclusive muon candidate yield from
  light-hadronic sources, as listed in
  Tables~\ref{tab:sigmaI}-\ref{tab:sigmaP} and discussed in
  Section~\ref{sec:syst}.}
\label{fig:ni_plusminussubtracted}
\end{figure*}

Many sources of systematic error on the yield of vertex-independent
muons, $N_{\mu}$, are common to the different components of the
inclusive muon candidate yield. In order to account for this we
rewrite Equation~\ref{eq:muonsources} (making the $p_T$ and $z_{vtx}$
dependencies implicit) as:
\begin{eqnarray}\label{eq:syserr0} 
N_{\mu} & = & N_I \times (N_{\mu}/N_I) \\ 
        & = & N_I \times ((N_I - N_D - N_P - N_B)/N_I) \nonumber \\
        & = & N_I \times (1 - R_D - R_P - R_B), \nonumber
\end{eqnarray}
where $R_j = N_j / N_I$ is the fraction of the inclusive muon
candidate yield attributed to the $j^{th}$ component. We can now write
the systematic error on $N_{\mu}$ as:
\begin{equation} \label{eq:syserr}
\sigma_{N_{\mu}} = \sqrt{(\sigma_{N_I}/N_I)^2 N^2_{\mu}  +
(\sigma^2_{R_D} + \sigma^2_{R_P} + \sigma^2_{R_B})N_I^2}
\end{equation}
$\sigma_{N_{\mu}}$, as determined below, is displayed in
Figures~\ref{fig:ni_plusminusdecomposed}
and~\ref{fig:ni_plusminussubtracted}. Note that the errors for
positives are significantly larger than for negatives. This is due to
the much larger relative contribution to positive inclusive muon
candidates from punchthrough hadrons, which is due to the relatively
small size of the $K^+$ nuclear interaction cross section.

Error sources contributing to $\sigma_{N_I}$ are quantified in
Table~\ref{tab:sigmaI}. Error sources contributing to $\sigma_{N_D}$
and $\sigma_{R_D}$ are quantified in Table~\ref{tab:sigmaD}. Error
sources contributing to $\sigma_{N_P}$ and $\sigma_{R_P}$ are
quantified in Table~\ref{tab:sigmaP}. Note that in these tables we
separately list errors that affect the overall normalization ($\sigma/
N^{\rm norm}$) and the shape of the $p_T$ spectrum
($\sigma/N^{p_T}$). The error on $R_B$ is taken to be 100\% of its
estimated value: $\sigma_{R_B} = N_B/N_I = 0.05 \times
N_D(z_{vtx}=0)/N_I$.

Values for $\sigma_{N_I}/N_I$ are displayed in
Figure~\ref{fig:ni_plusminus}.  Values for $\sigma_{R_D}$ and
$\sigma_{R_P}$ are displayed in Figures~\ref{fig:np_plusminus}
and~\ref{fig:nd_plusminus} respectively.  We insert
$\sigma_{N_I}/N_I$, $\sigma_{R_D}$, $\sigma_{R_P}$, and $\sigma_{R_B}$
into Equation~\ref{eq:syserr} as part of the final systematic error 
on $N_{\mu}$.

\begin{figure}[t]
\includegraphics[width=1.05\linewidth]{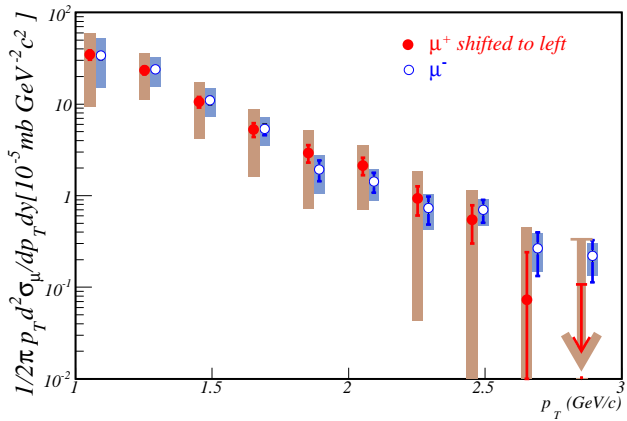}
\caption{$p_T$ spectrum of vertex-independent muons. Error bars
  indicate statistical errors. One point with unphysical (less than zero) extracted yield is shown as an arrow pointing down from the 90\% C.L.U.L. Shaded bands indicate systematic
  errors, as listed in
  Tables~\ref{tab:sigmaI}-\ref{tab:sigmaP} and discussed in
  Section~\ref{sec:syst}.}
\label{fig:nprompt_plusminus}
\end{figure}

To get the vertex-independent muon cross section, as defined in
Equation~\ref{eq:ptspectrum} and displayed in
Figures~\ref{fig:nprompt_plusminus} and~\ref{fig:pythia_fonll}, we
need to add in quadrature the errors on $N_{\mu}$,
$\sigma_{BBC}^{pp}$,
$\varepsilon_{BBC}^{c,\bar{c}\rightarrow\mu}$. The error on $N_{\mu}$
is obtained from the components above according to
Equation~\ref{eq:syserr}. As mentioned 


\clearpage

\noindent above, $\sigma_{BBC}^{pp}$ was
determined by a van der Meer scan to be 21.8\,mb, with an error of
9.6\%~\cite{ppg024}. We assign an error of 5\% to
$\varepsilon_{BBC}^{c,\bar{c}\rightarrow\mu}$, which was found to be
0.75 through two different methods~\cite{ppg017, ppg024}.


\begin{table*}[tbh]
\caption{Sources of systematic error on the calculation of $N_I$, the
  yield of inclusive muon candidates.}
\label{tab:sigmaI}
\begin{ruledtabular}\begin{tabular}{cccl}
Error Source & $\sigma/N^{\rm norm}$ & $\sigma/N^{p_T}$ & Comment \\ \hline
Momentum Scale             & 6.0\%   & $(p_T [{\rm GeV/c}] - 1) \times 1.3\%$ &  
Taken to be 100\% of the correction.\\
$\varepsilon_{\rm acc}$    & 10\%    & $(p_T [{\rm GeV/c}] - 1) \times 1.5\%$ & 
Taken to be 20\% of the correction.\\
$\varepsilon_{\rm rec}$    & 9.0\%   & 0       & 
Taken to be 25\% of the correction.\\
$\varepsilon_{\rm user}$   & 5.0\%   & $(p_T [{\rm GeV/c}] - 1) \times 5.0\%$ & 
Error on $\sigma^{\rm norm}$ taken to be 20\% of the correction.\\
&&& Error on $\sigma^{p_T}$ taken to be maximum of observed $p_T$
variation. \\
$\varepsilon_{\rm trig}$   & 4.7\%   & 0       & 
Taken to be the difference in $\varepsilon_{\rm trig}$ obtained with different procedures.\\ 
$\sigma_{N_I}/N_I$             & 16.3\%  & $(p_T [{\rm GeV/c}] - 1) \times
5.4\%$ & 
Add all rows in quadrature.
\\ 

\end{tabular}\end{ruledtabular}

\caption{Sources of systematic error on $R_D$, the ratio of free-decay muons to
  inclusive muon candidates, and $N_D$, the absolute yield of free-decay muons.}
\label{tab:sigmaD}
\begin{ruledtabular}\begin{tabular}{cccl}

Error Source & $\sigma/N^{\rm norm}$ & $\sigma/N^{p_T}$ & Comment \\ \hline

Decay flight path        & 5\%    & 0   & 
Variation due to shift of $\lambda_D$ by $\pm 3$\,cm.\\ 

$z_{vtx}$ fit range      & 3.3\%    & 0   & 
Variation seen using $|z_{vtx}|<20$\,cm, $|z_{vtx}|<40$\,cm.\\ 

Input hadron spectrum    & 0      & $(p_T [{\rm GeV/c}] - 1) \times 5.0\%$ & 
Spectrum normalized to data, so the only \\
&&& uncertainty here is in the $p_T$ dependence.\\

Decay normalization      & 7\%    & 0   & 
Statistical uncertainty in fit to observed free-decay muon yield.\\ 

$\sigma_{R_D}$           & 9.2\%   & $(p_T [{\rm GeV/c}] - 1) \times 5.0\%$ & 
Add all previous rows in quadrature.  \\ 

$\sigma_{N_I}/N_I$       & 16.3\%  & $(p_T [{\rm GeV/c}] - 1) \times 5.4\%$ & From Table~\ref{tab:sigmaI}. \\ 

$\sigma_{N_D}/N_D$       & 18.7\%  & $(p_T [{\rm GeV/c}] - 1) \times 7.4\%$ & 
Add previous two rows in quadrature.  \\ 

\end{tabular}\end{ruledtabular}

\caption{Sources of systematic error on $R_P$, the ratio of punchthrough
  hadrons to inclusive muon candidates, and $N_P$, the absolute yield of punchthrough hadrons.}
\label{tab:sigmaP}
\begin{ruledtabular}\begin{tabular}{cccl}

Error Source & $\sigma/N^{\rm norm}$ & $\sigma/N^{p_T}$ & Comment \\ \hline

Exponential absorption model    & 0       & 32\% &
Maximum fractional difference between $C^i(p_T)$ \\
&&& for FLUKA and GHEISHA, from Table~\ref{tab:h4ratio}.\\ 

$\varepsilon^3_{\rm scale}$     & 23\%    & 0 &
Dominated by ambiguity in the definition of which \\
&&& particles should be reconstructed.\\

$p$ and $\bar{p}$ contributions & 10\%    & 0 &
Variation with extreme assumptions on the $p$ and \\
&&&$\bar{p}$ nuclear interaction length.\\

$N_3^{stop}$ normalization      & 10\%     & 0 &
Statistical uncertainty in fit to observed gap 3 exclusive yield.\\ 

$\sigma_{R_P}$                  & 27\%    & 32\% &
Add all previous rows in quadrature. \\

$\sigma_{N_I}/N_I$              & 16.3\%  & $(p_T [{\rm GeV/c}] - 1) \times 5.4\%$ & From Table~\ref{tab:sigmaI}. \\ 

$\sigma_{N_P}/N_P$              & 31.5\%  & $\approx 32\%$ & 
Add previous two rows in quadrature.  \\ 

\end{tabular}\end{ruledtabular}
\end{table*}

\section{Charm Cross Section}
\label{sec:xsec}

The charm production cross section obtained from the yield of
vertex-independent muons (or from the yield of non-photonic electrons,
or $D$ mesons) is necessarily model dependent since we do not measure
the charm quarks directly. We use PYTHIA to convert our measurement of
the vertex-independent muon yield into an estimate of the differential
charm production cross section at forward rapidity,
$d\sigma_{c\bar{c}}/dy|_{y=1.6}$, in a procedure very similar to that
used in PHENIX measurements of charm production at
$y=0$~\cite{ppg011,ppg035,ppg037,kelly:2004,ppg056,butsyk:2005,ppg065}. We
use PYTHIA version 6.205 with parameters tuned to reproduce charm
production data at SPS and FNAL~\cite{Alves:1996} and single electron
data at the ISR~\cite{busser:1976, perez:1982, basile:1981}. Tuned
parameters are listed in Table~\ref{tab:pythia}. The meaning of each
parameter is more thoroughly defined in the PYTHIA
manual~\cite{PYTHIAmanual}.

\begin{table}[tbh]
\caption{Tuned PYTHIA parameters (default settings for this analysis) for
  determination of charm production cross section central value.}
\label{tab:pythia}

\begin{ruledtabular}\begin{tabular}{ccl} 
Parameter & Value & Meaning \\ \hline

MSEL & 4 & Heavy quark production every event \\
     &   & (gluon fusion + $q/\bar{q}$ annihilation). \\
MSTP(32)  & 4    & Hard scattering scale, $Q^2 = \hat{s}$. \\
MSTP(33)  & 1    & Use $K$-factor.      \\
MSTP(52)  & 2    & Use PDF libraries.   \\
MSTP(51)  & 4046 & Select CTEQ5L PDF libraries~\cite{Lai:2000}. \\ 
MSTP(91)  & 1    & Use Gaussian distribution for intrinsic $k_T$.   \\
PARP(31)  & 3.5  & $K$-factor.      \\
PARP(91)  & 1.5  & $<k_T>$ (GeV/c). \\
PARP(93)  & 5.0  & Maximum $k_T$ (GeV/c). \\
PMAS(4,1) & 1.25 & $m_c$ (GeV/c).   \\ 
$D^+/D^0$ & 0.32 & Default charm chemistry ratio. \\ 

\end{tabular}\end{ruledtabular}
\end{table} 

\begin{figure}[thb]
\includegraphics[width=1.05\linewidth]{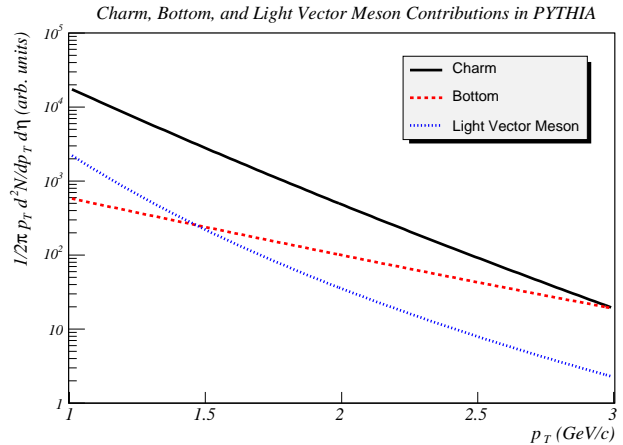}
\caption{PYTHIA calculation showing the major contributions to the
  vertex-independent muon $p_T$ spectrum. Solid, dashed and dotted
  lines show the yield from charm, bottom and short-decay length light
  vector mesons ($\rho, \omega, \phi$) respectively.}
\label{fig:prompt_pt}
\end{figure}

Vertex-independent muon sources, predicted by a PYTHIA simulation
using the same parameters (except that MSEL is set to 2 to generate
unbiased collisions), are listed in Table~\ref{tab:muonsources}. These
sources include decays of hadrons containing a heavy quark, and light
vector mesons with a decay length too short to be measured with the
existing experimental apparatus ($\rho, \omega, \phi$). Their $p_T$
spectra are shown in Figure~\ref{fig:prompt_pt}. Contributions from
quarkonium decays, Drell-Yan and $\tau$ lepton decays are
negligible. This shows that vertex-independent muon production in our
acceptance is dominated by muons from decay of charm hadrons, although
for $p_T > 2.5$\,GeV/c the contribution from decays of hadrons
containing a bottom quark is starting to become important.

\begin{table}[tbh]
\caption{Percentage contribution of different sources of
  vertex-independent muons within our acceptance ($1<p_T<3$\,GeV/c and
  $1.5<|\eta|<1.8$), from PYTHIA, with parameters listed in
  Table~\ref{tab:pythia} (except that ${\rm MSEL}=2$ to generate
  minimum bias collisions).}
\label{tab:muonsources}
\begin{ruledtabular}\begin{tabular}{ll}
Source & Contribution \\ \hline
Open charm & 84.6\% \\
Open bottom & 6.9\% \\
$\rho, \omega, \phi$ & 8.1\% \\
Quarkonia & $< 0.1$\% \\
Drell-Yan & $< 0.1$\% \\
$\tau$ leptons & 0.4\% \\
\end{tabular}\end{ruledtabular}
\end{table}

\begin{figure}[thb]
\includegraphics[width=1.05\linewidth]{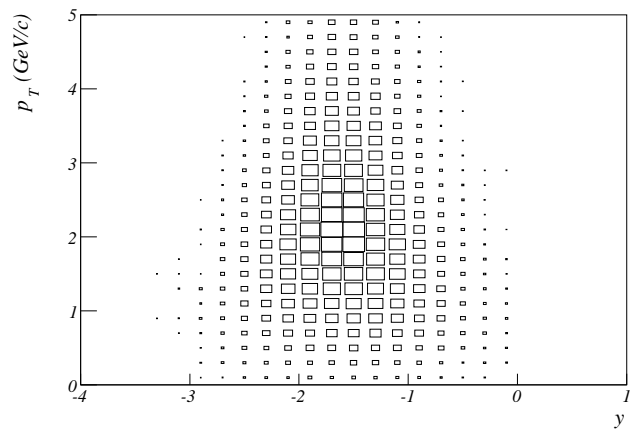}
\caption{PYTHIA results for the $p_T$ vs.~$y$ distribution (linear
  $z$-scale) of charm quarks that produce a muon in the PHENIX
  acceptance.}
\label{fig:charm_pt_y}
\end{figure}

This simulation also gives the distribution of charm quarks ($p_T$
vs.~$y$) that produce a muon in our acceptance, as shown in
Figure~\ref{fig:charm_pt_y}. This demonstrates that the
vertex-independent muons we measure sample charm quarks down to $p_T
\approx 1$\,GeV/c, over a narrow rapidity slice centered at $y=1.6$.

\begin{figure}[thb]
\includegraphics[width=0.8\linewidth]{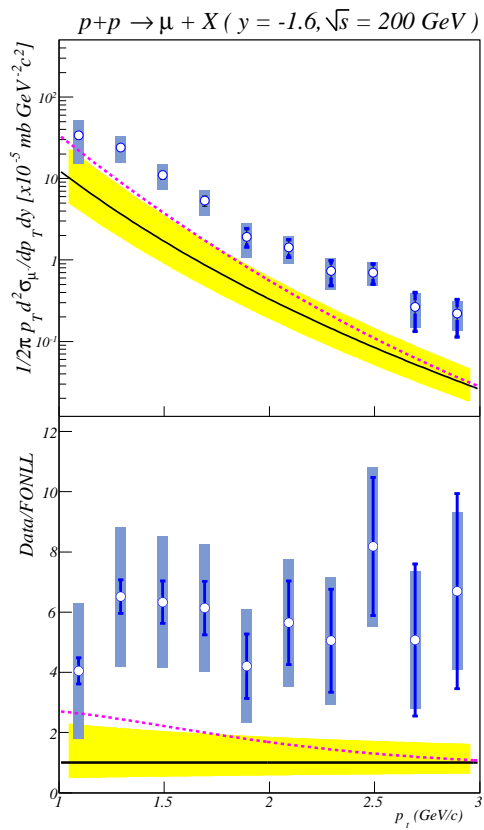}
\caption{The top panel shows the measured $p_T$ spectrum of
  vertex-independent negative muons from
  Figure~\ref{fig:nprompt_plusminus}, the PYTHIA prediction using
  settings listed in Table~\ref{tab:pythia} without scaling the charm
  contribution (dotted line), and a FONLL calculation (solid line with
  systematic error band)~\cite{cacciari:2005,Cacciari_pc}. The bottom
  panel shows the ratio of the measured spectrum to the FONLL
  calculation with statistical (error bars) and systematic (bands)
  uncertainties on the data, as well as the theoretical uncertainty
  (shaded band around 1). The dashed line shows the PYTHIA/FONLL
  ratio.}
\label{fig:pythia_fonll}
\end{figure}

Figure~\ref{fig:pythia_fonll} shows a comparison of the measured
vertex-independent negative muon spectrum (from
Figure~\ref{fig:nprompt_plusminus}) to the prediction of this default
PYTHIA simulation and to a FONLL calculation~\cite{cacciari:2005,
Cacciari_pc}. One can see that the measured values significantly
exceed both predictions. The spectrum also appears to be somewhat
harder than the PYTHIA spectrum with the parameters listed in
Table~\ref{tab:pythia}.

We scale the charm (only) contribution to the PYTHIA
vertex-independent muon $p_T$ spectrum such that the total spectrum
(including the small contributions from open bottom and vector mesons)
matches the central values of the measured vertex-independent negative
muon spectrum. Only statistical errors are used in the fit. Note,
larger systematic errors for the positive muon spectrum preclude a
significant measurement for that charge sign. We multiply the scale
factor from the fit (2.27) by the PYTHIA value for the charm
production cross section, $d\sigma_{c\bar{c}}/dy|^{PYTHIA}_{y=1.6}$
(0.107\,mb), to obtain $d\sigma_{c\bar{c}}/dy|^{PHENIX}_{y=1.6} =
0.243 \pm 0.013 {\rm (stat.)}$\,mb.
 
We distinguish between two different sources of systematic uncertainty
on the extraction of the charm cross section: 1) uncertainty in the
PYTHIA calculation, and 2) uncertainty in the data, which is largely
independent of PYTHIA.

We determined the uncertainty in the data ($\pm 43\%$) by refitting
PYTHIA to the data at the minimum and maximum of the 1$\sigma$
systematic error band.

We determined the uncertainty in the PYTHIA calculation with a
systematic study in which we varied simulation parameters, extracted
the new simulated vertex-independent negative muon spectrum,
normalized to the measured spectrum, and extracted
$d\sigma_{c\bar{c}}/dy|^{PHENIX}_{y=1.6}$ for the modified parameter
sets. We varied PDF libraries, the hard scattering scale, the charm
quark mass, the intrinsic $k_T$ value, the $D^+/D^0$ ratio, charm
production mechanism selections, and open bottom and vector meson
scaling assumptions. The parameter sets used and the results of this
study are summarized in Table~\ref{tab:cross}.

\begin{table*}[tbh]
\caption{Results for PYTHIA simulations with different parameter sets
used to explore the systematic error on the charm cross section due to
model uncertainties. The top of the table details the different
parameter sets tested. Unless otherwise noted, parameters are the same
as those listed in Table~\ref{tab:pythia}.  The bottom of the table
gives the results for different simulations: The $1^{st}$
column identifies the simulation; the $2^{nd}$ column gives the total
charm production cross section given the chosen PYTHIA parameter set;
the $3^{rd}$ column gives the differential charm production cross
section at $y=1.6$; the $4^{th}$ column gives the normalization factor
needed to fit the PHENIX data; the $5^{th}$ column gives the
differential charm production cross section at $y=1.6$ for PHENIX data
(the product of the $3^{rd}$ and $4^{th}$ columns); the $6^{th}$
column gives the fractional difference between the results for each
simulation compared to the simulation with the default PYTHIA
parameter set; the last column gives the ratio
$d\sigma_{c\bar{c}}/dy|^{PYTHIA}_{y=1.6}/d\sigma_{c\bar{c}}/dy|^{PYTHIA}_{y=0}$. }
\label{tab:cross}

\begin{ruledtabular}\begin{tabular}{ccccccc}

Case & \multicolumn{6}{c}{PYTHIA Settings} \\ \hline

1  & \multicolumn{6}{l}{Default settings, see Table~\ref{tab:pythia}.} \\ 
2a & \multicolumn{6}{l}{${\rm MSTP(51) = 4032}$, CTEQ4L PDF libraries~\cite{cteq4l}.} \\
2b & \multicolumn{6}{l}{${\rm MSTP(51) = 5005}$, GRV94LO PDF libraries~\cite{grv94lo}.} \\
2c & \multicolumn{6}{l}{${\rm MSTP(51) = 5012}$, GRV98LO PDF libraries~\cite{grv98lo}.} \\
2d & \multicolumn{6}{l}{${\rm MSTP(51) = 3072}$, MRST (c-g) PDF libraries~\cite{mrstcg}.} \\
3a & \multicolumn{6}{l}{${\rm MSTP(32) = 1}, Q^2 = 2\hat{s}\hat{t}\hat{u} /
(\hat{s}^2 + \hat{t}^2 + \hat{u}^2)$.} \\
3b & \multicolumn{6}{l}{${\rm MSTP(32) = 2}, Q^2 = p_T^2 + (m_3^2 + m_4^2)/2$.} \\
3c & \multicolumn{6}{l}{${\rm MSTP(32) = 3}, Q^2 = \min(-\hat{t}, -\hat{u})$.} \\
3d & \multicolumn{6}{l}{${\rm MSTP(32) = 5}, Q^2 = -\hat{t}$.} \\
4a & \multicolumn{6}{l}{${\rm PMAS(4,1)} = m_c = 1.15$\,GeV/c.} \\
4b & \multicolumn{6}{l}{${\rm PMAS(4,1)} = m_c = 1.35$\,GeV/c.} \\
4c & \multicolumn{6}{l}{${\rm PARP(91)} = <k_T> = 0.3$\,GeV/c.} \\
4d & \multicolumn{6}{l}{${\rm PARP(91)} = <k_T> = 3.0$\,GeV/c.} \\
4e & \multicolumn{6}{l}{${\rm MSTP(68) = 2}$, Maximum virtuality scale and matrix element matching scheme.} \\
   & \multicolumn{6}{l}{${\rm PARP(67) = 4}$, Multiplicative factor applied to hard scattering scale.} \\
5a & \multicolumn{6}{l}{${\rm PARP(31)} = K$-factor $ = 1$,} \\
   & \multicolumn{6}{l}{${\rm MSEL} = 1$, Hard scattering enabled.} \\
5b & \multicolumn{6}{l}{${\rm PARP(31)} = K$-factor $ = 1$,} \\ 
   & \multicolumn{6}{l}{${\rm MSEL} = 1$, Hard scattering enabled,}\\
   & \multicolumn{6}{l}{All other parameters untuned.} \\
6  & \multicolumn{6}{l}{$D^+/D^0 = 0.45$~\cite{STARDchemistry}.} \\
7  & \multicolumn{6}{l}{Open bottom and vector mesons scale with charm.} \\ 

Case                                               & 
$\sigma^{PYTHIA}_{c\bar{c}}$                       & 
$d\sigma_{c\bar{c}}/dy|^{PYTHIA}_{y=1.6}$          & 
Normalization                                      & 
$d\sigma_{c\bar{c}}/dy|^{PHENIX}_{y=1.6}$          & 
$\Delta d\sigma_{c\bar{c}}/dy|^{PHENIX}_{y=1.6}$   & 
$d\sigma_{c\bar{c}}/dy|^{PYTHIA}_{y=1.6}/$         \\ 

        & 
(mb)    & 
(mb)    & 
to Data & 
(mb)    & 
(\%)    & 
$d\sigma_{c\bar{c}}/dy|^{PYTHIA}_{y=0}$         \\ \hline

1   & 0.658 & 0.107 & 2.27 & 0.243 &   --  & 0.67\\ 
2a  & 0.691 & 0.111 & 2.10 & 0.232 &  -4.5 & 0.69\\
2b  & 0.698 & 0.112 & 2.09 & 0.233 &  -3.9 & 0.71\\
2c  & 0.669 & 0.109 & 2.18 & 0.238 &  -1.7 & 0.73\\
2d  & 0.551 & 0.088 & 2.67 & 0.236 &  -2.9 & 0.71\\
3a  & 1.520 & 0.243 & 1.12 & 0.271 &  11.8 & 0.84\\
3b  & 0.863 & 0.139 & 1.63 & 0.226 &  -6.7 & 0.71\\
3c  & 1.501 & 0.242 & 1.11 & 0.267 &  10.2 & 0.84\\
3d  & 1.104 & 0.178 & 1.45 & 0.258 &   6.4 & 0.78\\
4a  & 0.905 & 0.145 & 1.73 & 0.252 &   3.7 & 0.67\\
4b  & 0.487 & 0.078 & 2.91 & 0.226 &  -6.7 & 0.64\\
4c  & 0.658 & 0.104 & 2.81 & 0.292 &  20.4 & 0.66\\
4d  & 0.658 & 0.104 & 1.50 & 0.156 & -35.8 & 0.63\\
4e  & 0.658 & 0.106 & 2.09 & 0.220 &  -9.2 & 0.63\\
5a  & 0.435 & 0.068 & 3.91 & 0.266 &   9.4 & 0.80\\
5b  & 0.385 & 0.058 & 4.67 & 0.271 &  11.7 & 0.79\\
6   & 0.658 & 0.107 & 2.38 & 0.255 &   5.0 & 0.67\\   
7   & 0.658 & 0.107 & 2.20 & 0.236 &  -2.9 & 0.67\\

\end{tabular}\end{ruledtabular}
\end{table*}

The PYTHIA charm cross section varies substantially
($\Delta(d\sigma_{c\bar{c}}/dy)|^{PYTHIA}_{y=1.6}) \approx 4$) for the
chosen parameter sets. However, the extracted experimental charm cross
section is relatively stable
($\Delta(d\sigma_{c\bar{c}}/dy)|^{PHENIX}_{y=1.6} < 0.36$). This is
due to the fact that the parameter set changes have relatively minor
effects on the shape of the predicted vertex-independent muon $p_T$
spectrum, and we obtain the experimental charm cross section by
normalizing the PYTHIA charm cross section by the ratio of the
measured and predicted muon $p_T$ spectra.

\begin{figure*}[thb]
\includegraphics[width=0.5\linewidth]{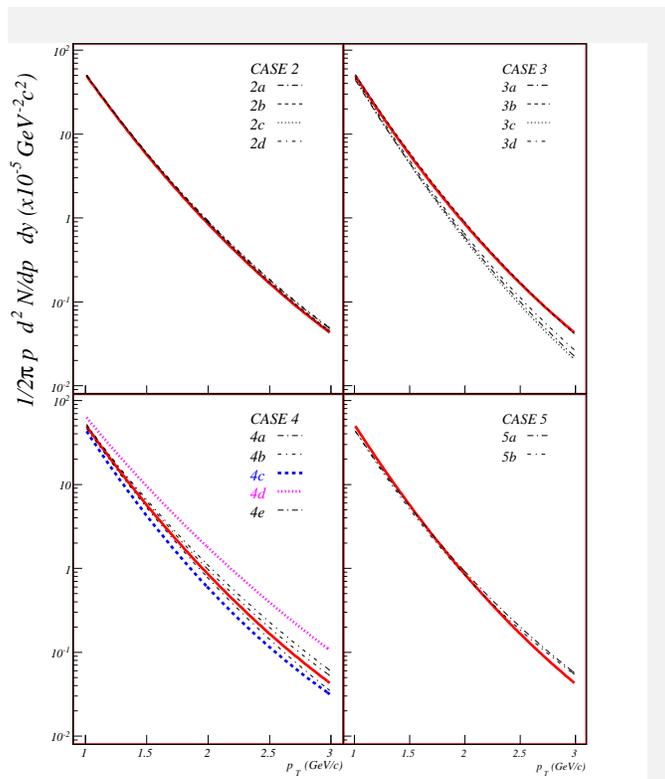}
\caption{(Color online) PYTHIA results with different parameter sets
  for the negative vertex-independent muon $p_T$ spectrum per event in
  which a $c\bar{c}$ pair is created. The solid line in each panel
  shows the result when using default settings listed in
  Table~\ref{tab:pythia}. Legends indicate the correspondance between
  line style and the simulation case label. Parameter sets for each
  case label are given in Table~\ref{tab:cross}.}
\label{fig:PYTHIA_muon_shapes}
\end{figure*}

One way to visualize this is to plot (see
Figure~\ref{fig:PYTHIA_muon_shapes}) the vertex-independent muon yield
in our acceptance {\em per event in which a $c\bar{c}$ pair is
created} for the different PYTHIA parameter sets. Due to our
procedure, parameter sets which give similar vertex-independent muon
yields per $c\bar{c}$ event in the low $p_T$ region (which dominates
the fit) will necessarily give similar values for
$d\sigma_{c\bar{c}}/dy|^{PHENIX}_{y=1.6}$, whatever the PYTHIA charm
cross section is.

The largest variation in the predicted muon yield at $p_T = 1$\,GeV/c
per $c\bar{c}$ event is seen for simulations in which the intrinsic
$k_T$ is varied from its default value ($<k_T> = 1.5$\,GeV/c) to the
value expected from arguments based on Fermi momentum (case 4c, $<k_T>
= 0.3$\,GeV/c), or to a value which best reproduces the measured
spectrum at higher $p_T$ (case 4d, $<k_T> = 3.0$\,GeV/c). These
parameter sets also result in the largest variation in
$d\sigma_{c\bar{c}}/dy|^{PHENIX}_{y=1.6}$, as shown in
Table~\ref{tab:cross}. We use the cross section values obtained in
this pair of simulations to define the systematic uncertainty in our
measurement due to the uncertainty in our PYTHIA calculation. This
gives us our final answer: $d\sigma_{c\bar{c}}/dy|_{y=1.6} = 0.243 \pm
0.013 {\rm (stat.)}  \pm 0.105 {\rm
(data~syst.)}~^{+0.049}_{-0.087}{\rm (PYTHIA~syst.)}$\,mb.

\begin{figure}[thb]
\includegraphics[width=0.7\linewidth]{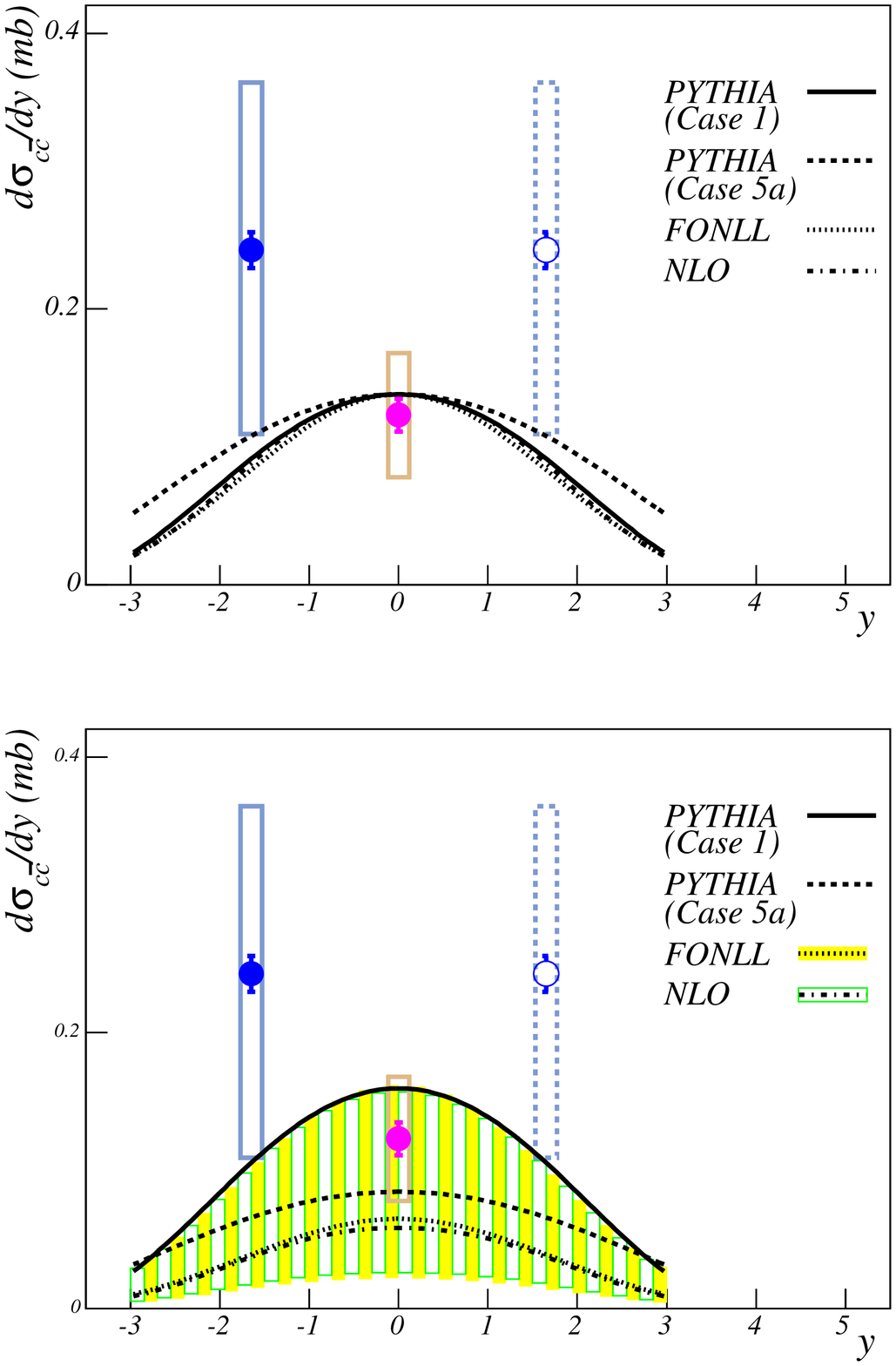}
\caption{(Color online) Comparisons of measured charm rapidity
distributions, $d\sigma_{c\bar{c}}/dy$ vs.~$y$, to theoretical
predictions. Data points at $y=\pm1.6$ are from this analysis (the
point at $y=1.6$ is reflected through $y=0$). The point at $y=0$ is
the PHENIX measurement of charm through semileptonic decay to
electrons~\cite{ppg065}. Error bars on the data points indicate
statistical uncertainties and boxes indicate systematic
uncertainties. The top panel shows rapidity spectra from two PYTHIA
parameter sets (see Table~\ref{tab:cross} for details),
FONLL~\cite{cacciari:2005,Cacciari_pc}, and an NLO
calculation~\cite{Vogt_pc}. The PYTHIA curve with the default
parameter set (Case 1) was fit to the two PHENIX data points with
statistical and systematic errors added in quadrature. All other theory
curves were normalized so that they are equal at $y=0$ to allow shape
comparisons. The bottom panel shows the theory curves
unnormalized. Theoretical uncertainties associated with the FONLL and
NLO calculations are indicated with shaded bands.}

\label{fig:dsdy_vs_y}
\end{figure}

Figure~\ref{fig:dsdy_vs_y} shows the PHENIX charm rapidity
spectrum. The result of this analysis (mirrored about $y=0$ since this
is a symmetric collision system) is plotted along with the 
result for $d\sigma_{c\bar{c}}/dy|_{y=0}$~\cite{ppg065}. In order to
compare with the data at $y=0$ the systematic uncertainty on the data
from this analysis is shown as the quadrature sum of the two sources
of systematic uncertainty described above (data and
PYTHIA). Theoretical curves from PYTHIA (case 1 and case 5a),
FONLL~\cite{cacciari:2005, Cacciari_pc}, and an NLO calculation from
Vogt~\cite{Vogt_pc} are also displayed.

In the top panel of the figure PYTHIA with the default parameter set
(Case 1) is fit to the two PHENIX points with statistical and
systematic errors added in quadrature. Other theory curves are
normalized so that they are equal at $y=0$ in order to allow shape
comparisons. As shown in Table~\ref{tab:cross}, different PYTHIA
parameter sets differ in the predicted ratio
$d\sigma_{c\bar{c}}/dy|^{PYTHIA}_{y=1.6}/d\sigma_{c\bar{c}}/dy|^{PYTHIA}_{y=0}$
by $>30\%$. Unfortunately, current systematic error bars preclude any
conclusions about the charm production rapidity shape.

In the bottom panel of the figure the theory curves are unnormalized
to allow an absolute comparison. The quoted theoretical uncertainty
bands for the FONLL and NLO calculations are also shown. We note that,
although our data are above the FONLL prediction, the error bars
touch. This is in contrast to the situation for the vertex-independent
muon cross section, shown in Figure~\ref{fig:pythia_fonll}, where the
data are significantly above the prediction. The larger disagreement
in the vertex-independent muon cross section is presumably due to
different treatment of the fragmentation process in PYTHIA and
FONLL~\cite{cacciari:2005,Vogt_pc, Cacciari_pc}.

\section{Conclusion and Outlook}
\label{sec:conclusion}

We have measured muon production at forward rapidity ($1.5 \le |\eta|
\le 1.8$), in the range $1 < p_T < 3$\,GeV/c, in $\sqrt{s} = 200$\,GeV
$p+p$ collisions at RHIC. We determined and subtracted the
contribution from light hadron sources ($\pi, K, p$) to obtain the
vertex-independent muon yield which, for the $p_T$ range measured in
this analysis, and in the absence of new physics, arises dominantly
from the decay of $D$ mesons. We normalized the PYTHIA muon spectrum
resulting from the production of charm quarks to obtain the
differential cross section for charm production at forward rapidity:
$d\sigma_{c\bar{c}}/dy|_{y=1.6} = 0.243 \pm 0.013 {\rm (stat.)}  \pm
0.105 {\rm (data~syst.)}~^{+0.049}_{-0.087}{\rm (PYTHIA~syst.)}$\,mb.
This is compatible with PHENIX charm measurement at $y=0$, although
even further above predictions from PYTHIA and FONLL. Large systematic
uncertainties in the current measurement preclude statements about the
rapidity dependence of the charm cross section.

The systematic uncertainty in the data is dominated by uncertainty on
the determination of the fractional contribution of decay muons. This
will be improved with higher statistics data sets (already collected)
which will allow better measurements of the $z_{vtx}$ dependence of
particle production. Final results for identified particle $p_T$
distributions in $p+p$ collisions by BRAHMS will also be invaluable
for improving the input to our hadron generator. The systematic
uncertainty in PYTHIA is dominated by differences observed when the
intrinsic $<k_T>$ is varied. In order to reduce this uncertainty we
need to reduce the allowed parameter space by improving the
measurement of the high $p_T$ portion of the vertex-independent muon
spectrum, where the error is dominated by the uncertainty in the yield
of punchthrough hadrons. Data sets (already collected) with higher
statistics, and with hadrons stopping in MuID gap 2, will allow a
completely data-driven approach to the calculation of the punchthrough
yield. This will eliminate the reliance on hadronic interaction
simulation packages, differences in which are the largest source of
systematic error at high $p_T$. Analogous measurements are also being
carried out for $d+Au$, $Cu+Cu$, and $Au+Au$~\cite{xrwang_PANIC}
collisions at $\sqrt{s_{NN}} = 200$\,GeV. These will allow
determination of the magnitude of nuclear modification effects on
charm production at forward rapidity.


\section{Acknowledgements}   

We thank the staff of the Collider-Accelerator and Physics
Departments at Brookhaven National Laboratory and the staff
of the other PHENIX participating institutions for their
vital contributions.  We acknowledge support from the
Department of Energy, Office of Science, Nuclear Physics
Division, the National Science Foundation, Abilene Christian
University Research Council, Research Foundation of SUNY, and
Dean of the College of Arts and Sciences, Vanderbilt
University (U.S.A), Ministry of Education, Culture, Sports,
Science, and Technology and the Japan Society for the
Promotion of Science (Japan), Conselho Nacional de
Desenvolvimento Cient\'{\i}fico e Tecnol{\'o}gico and Funda\c
c{\~a}o de Amparo {\`a} Pesquisa do Estado de S{\~a}o Paulo
(Brazil), Natural Science Foundation of China (People's
Republic of China), Centre National de la Recherche
Scientifique, Commissariat {\`a} l'{\'E}nergie Atomique,
Institut National de Physique Nucl{\'e}aire et de Physique
des Particules, and Institut National de Physique
Nucl{\'e}aire et de Physique des Particules, (France),
Bundesministerium f\"ur Bildung und Forschung, Deutscher
Akademischer Austausch Dienst, and Alexander von Humboldt
Stiftung (Germany), Hungarian National Science Fund, OTKA
(Hungary), Department of Atomic Energy and Department of
Science and Technology (India), Israel Science Foundation
(Israel), Korea Research Foundation and 
Korea Science and Engineering Foundation (Korea), 
Russian Ministry of Industry, Science
and Tekhnologies, Russian Academy of Science, Russian
Ministry of Atomic Energy (Russia), VR and the Wallenberg
Foundation (Sweden), the U.S. Civilian Research and
Development Foundation for the Independent States of the
Former Soviet Union, the US-Hungarian NSF-OTKA-MTA, the
US-Israel Binational Science Foundation, and the 5th European
Union TMR Marie-Curie Programme.

\clearpage


\end{document}